\documentclass[aps,prx,reprint,twocolumn,floatfix,superscriptaddress,amsmath,amssymb,amsfonts,nofootinbib]{revtex4-2}

\usepackage{url}
\usepackage{bm}
\usepackage{graphicx}
\usepackage{amsmath}
\usepackage{amstext}
\usepackage{amssymb}
\usepackage{amsfonts}
\usepackage{amsbsy}
\usepackage{verbatim}
\usepackage{color}
\usepackage[colorlinks=true, urlcolor=blue, linkcolor=blue, citecolor=blue, pdftex]{hyperref}
\usepackage{floatrow}
\usepackage{float}
\usepackage{tikz}
\usepackage{gensymb}
\usepackage{textcomp}
\usepackage{enumitem}
\usepackage[version=3]{mhchem}
\usepackage{afterpage}
\usepackage{pbox}
\usepackage{dsfont}
\usepackage{siunitx}
\usepackage{stackengine,wasysym,scalerel}
\usepackage{latexsym}
\usepackage{cancel}
\usepackage{comment}
\usepackage{array, multirow, bigdelim, makecell, booktabs}
\usepackage[misc]{ifsym}
\usepackage[capitalise]{cleveref}
\usepackage{sidecap}
\usepackage{xcolor}
\usepackage{lipsum}

\usepackage{color}
\usepackage[colorlinks=true, urlcolor=blue, linkcolor=blue, citecolor=blue, pdftex]{hyperref}

\usepackage[normalem]{ulem}

\definecolor{darkblue}{HTML}{004D6B}
\definecolor{darkred}{HTML}{8c1515}
\definecolor{darkgreen}{HTML}{006400}
\usepackage{hyperref}
\hypersetup{
	colorlinks=true,
	urlcolor=blue,
	citecolor=blue,
	linkcolor=blue,
	breaklinks
}

\graphicspath{{figures/}}
\allowdisplaybreaks

\def\h{\hbar}

\def\ka{\kappa}

\def\r{{\mathbf{r}}}

\newcommand{\pd}[2]{\frac{\partial #1}{\partial #2}}

\newcommand{\bra}[1]{\langle #1|}
\newcommand{\ket}[1]{|#1\rangle}

\begin{document}
\title{High-spin magnetic ground states of neutral dopant clusters in semiconductors}

\author{Rhine Samajdar}
\thanks{These two authors contributed equally.}
\affiliation{Department of Electrical and Computer Engineering, Princeton University, Princeton, NJ 08544, USA}
\affiliation{Department of Physics, Princeton University, Princeton, NJ 08544, USA}

\author{Haonan Zhou}
\thanks{These two authors contributed equally.}
\affiliation{Department of Physics, Princeton University, Princeton, NJ 08544, USA}
\affiliation{Tower Research Capital, New York, NY 10271, USA}

\author{R. N. Bhatt}
\affiliation{Department of Electrical and Computer Engineering, Princeton University, Princeton, NJ 08544, USA}
\affiliation{Department of Physics, Princeton University, Princeton, NJ 08544, USA}

\begin{abstract}
High-spin states hold significant promise for classical and quantum information storage and emerging magnetic memory technologies.
Here, we present a systematic framework for engineering such high-spin magnetic states in dopant clusters formed from substitutional impurities in semiconductors. In single-valley materials such as gallium arsenide, impurity states are hydrogenic and exchange interactions generally favor low-spin configurations, except in special geometries. In contrast, multivalley semiconductors exhibit oscillatory form factors in their exchange couplings, enabling the controlled suppression of selected hopping processes and exchange couplings. Exploiting this feature, we demonstrate how carefully arranged impurities in aluminum arsenide, germanium, and silicon can stabilize ground states with a net spin that scale extensively with system size. Within effective mass theory and the tight-binding approximation for hopping, we construct explicit examples ranging from finite clusters to extended lattices and fractal-like tilings. In two dimensions, we identify several favorable dopant geometries supporting a net spin equal to around half of the fully polarized value in the thermodynamic limit, including one which achieves over $70\%$ polarization. Our results provide a general design principle for harnessing valley degeneracy in semiconductors to construct robust high-spin states and outline a pathway for their experimental realization via precision implantation of dopants.
\end{abstract}
\date{\today}

\maketitle

\section{Introduction}

The stabilization and control of high-spin magnetic states presents new opportunities in the design of next-generation information technologies \cite{vzutic2004spintronics,bader2010spintronics}. On the \textit{atomic} scale, single-molecule magnets---constructed from large rings and metal-oxide clusters of atoms such as Mn, Dy, and Tb---have been shown to exhibit collective bistable magnetic behavior, which holds promise for information storage at the level of individual molecules \cite{friedman1996macroscopic}. Much effort has been devoted to enhancing the blocking temperature of such clusters through chemical design strategies, with the ultimate aim of realizing stable magnetic memories suitable for real-world devices \cite{hill2025making}. In the process, studying the delicate interplay of molecular large-spin states with thermal and quantum fluctuations has uncovered pathways toward not only denser classical magnetic storage but also controllable spin qubits, where these discrete spin states serve as addressable quantum resources \cite{ardavan2007will,bayliss2020optically}.\,Understanding and realizing robust high-spin states is therefore, in addition to being a technologically relevant problem, of fundamental importance to condensed matter and quantum information science.
Inspired by the bottom-up, atom-by-atom synthesis of molecular magnets, here, we propose a complementary route to assembling high-spin clusters on the \textit{mesoscopic} scale, using ensembles of substitutional impurities in semiconductors.

In particular, $n$-doped semiconductors with low concentrations of shallow dopants (well below the insulator--metal transition) can be modeled as systems of hydrogenic centers, with binding energies and effective Bohr radii $a^{*}_B$ that depend only on the host material's dielectric constant and effective mass in the conduction band \cite{Kohn,Ramdas, Thomas}. When the typical donor--donor distance $d\!\gg \!a^{*}_B$,  their magnetic properties are quantitatively captured by an effective Heisenberg Hamiltonian with pairwise exchange interactions \cite{Herring,bhatt1980clustering,bhatt1981scaling,bhatt1982scaling,sarachik1986scaling}. In the case of uncompensated systems (one electron per dopant), the natural exchange interaction is antiferromagnetic, which generally favors low-spin ground states for most cluster geometries.

However, because the exchange interaction decays exponentially with distance, there exist special geometries where high-spin ground states can arise \cite{Nielsen}. A notable example is a wheel-shaped cluster, where a central site is surrounded by several equidistant outer sites (see Sec.~\ref{sec:direct}). In this configuration, the exchange between the central site and each outer site dominates over the exchange between outer sites, leading to an enhanced ground state spin. Unfortunately, this mechanism is limited: once the number of outer sites grows too large, exchange among the outer sites becomes significant and drives the system back to a low-spin state. In two dimensions, this occurs for six outer sites arranged in a hexagonal geometry, such that the maximum achievable ground-state spin is $S=2$ for five outer sites. 

One strategy to circumvent this limitation is to consider compensated (or ``anticompensated'') systems, where the number of electrons is smaller (or larger) than the number of dopants. In the dilute limit, such systems can be mapped onto a Hubbard model \cite{hubbard1964electron}, with an effective  ratio of the interaction ($U$) to the hopping ($t$) that grows very large as the doping decreases. In the limit of $U/t \rightarrow \infty$, a rigorous result by \citet{nagaoka1966ferromagnetism} asserts that the addition or removal of a single electron relative to the half-filled case drives the system into a ferromagnetic (maximal-spin) ground state \cite{thouless1965exchange,tasaki1989extension,tian1990simplified}. While such Nagaoka-like ferromagnetism has been widely investigated with ultracold atoms in optical lattices---both theoretically \cite{samajdar2023nagaoka,sb2,schlomer2023,li2024nagaoka} and experimentally \cite{xu2023frustration,lebrat2023observation,prichard2023}---for conventional doped semiconductors, positional disorder effects appear to suppress this mechanism \cite{nielsen2007nanoscale}. More recently, however, the experimental realization of ordered lattice geometries in large arrays of semiconductor quantum dots \cite{Vandersypen, wang2022experimental, kiczynski2022engineering} has opened the door to implementing diverse theoretical proposals for large-spin ground states \cite{buterakos2023certain,buterakos2023magnetic,dieplinger2024itinerant}.  

In this work, we adopt a different approach: we focus on neutral clusters (one electron per site) in many-valley semiconductors. In such systems, the electronic ground-state wavefunction is a symmetric superposition of contributions from all equivalent conduction band minima. This structure modifies the exchange interaction: it remains exponentially decaying, as in the purely hydrogenic case, but acquires an additional form factor that oscillates between zero and one \cite{Andres,koiller2001exchange,koiller2002strain,wellard2003}. By carefully placing dopant atoms on the host semiconductor lattice, it is possible to exploit this form factor to stabilize cluster ground states with significantly larger spin.  This offers a particularly promising route for the construction of high-spin ground states, as established experimental techniques already enable the  placement of impurities at precise locations within semiconductor lattices \cite{Simmons, tranter2024machine}. Given the current state of the field, we concentrate on two-dimensional (or quasi-2D) structures in most of this paper, though a generalization to three-dimensional clusters or superlattices is straightforward.

This paper is organized as follows.  
First, in Sec.~\ref{sec:direct}, we begin with the simple case of purely hydrogenic clusters---as realized in direct bandgap semiconductors with a single valley, such as gallium arsenide---and examine their magnetic properties. Then, in Sec.~\ref{sec:math}, we turn to systems with degenerate conduction band minima located away from the Brillouin zone center. Here, the effective masses are anisotropic and impurity-state wavefunctions deviate from the purely hydrogenic form, as mentioned above. Although this makes the analysis more challenging, we show how it also introduces additional features that can be exploited to engineer clusters with ground states of exceptionally high spin. We develop this more complex scenario in detail in Sec.~\ref{sec:indirect}, where, within effective mass theory \cite{Kohn}, we demonstrate how to design dopant clusters with arbitrarily large spin both in semiconductors whose conduction band minima lie on the Brillouin zone boundary (e.g., germanium and aluminum arsenide), and in materials where the minima are located within the Brillouin zone interior (e.g., silicon). Finally, we provide a summary of our main results  in Sec.~\ref{sec:end} and discuss prospects for extending this construction to higher spatial dimensions.

\section{Direct bandgap semiconductors}
\label{sec:direct}

\begin{figure}[t]
\begin{center}
\includegraphics[width=\linewidth]{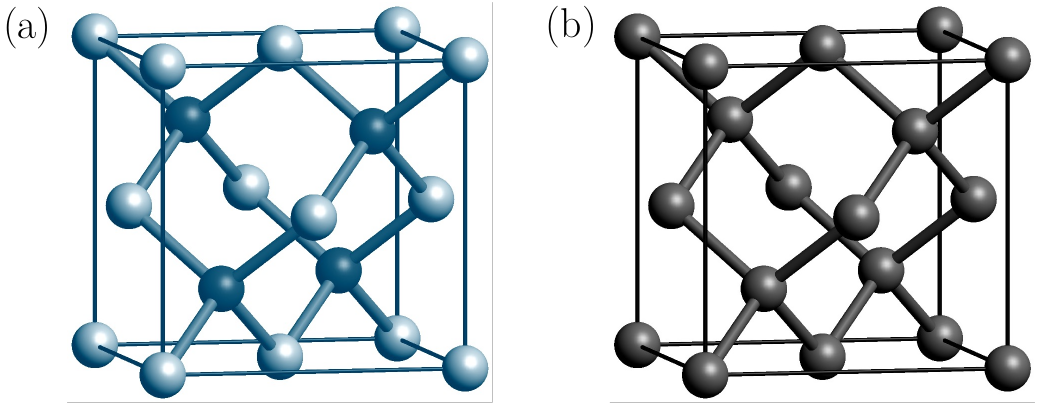}
\caption{Unit cells of (a) zincblende and (b) diamond cubic crystal structures. Both consist of two interpenetrating face-centered cubic lattices; however, zincblende  has two distinct atomic species, shown here in light and dark blue.}
\label{fig:directLattice}
\end{center}
\end{figure}

When a host semiconductor has a nondegenerate and isotropic conduction band minimum at the Brillouin zone center, the eigenenergies and envelope wavefunctions of shallow impurity states are hydrogenic in character \cite{Kittel1}.
To develop a quantitative picture, let us consider the simplified case where a lattice atom is replaced by an impurity atom whose atomic number exceeds that of the host by one. Removing a carrier electron from this impurity leaves behind a positively charged ion embedded in the neutral lattice. The resulting ion polarizes the surrounding semiconductor, and at large distances, the Coulomb interaction is given by
$
V(\r) = -{e^2}/{\ka r},
$
where $\kappa$ denotes the static dielectric constant of the crystal.  

For shallow impurities in particular, the bound carrier occupies a spatially extended orbital with a small ionization energy. In this case, the long-range potential $V(\r)$ provides a good approximation for the impurity contribution. Consequently, the carrier wavefunction is governed by the Schr\"odinger equation
\begin{equation} \label{eq:qualEq}
\left( -\frac{\h^2}{2m^*} \nabla^2 - \frac{e^2}{\ka r} - E \right) F(\r) = 0,
\end{equation}
where $m^*$ is the effective mass. The solutions of Eq.~\eqref{eq:qualEq} are formally identical to the hydrogen atom, except that the fundamental energy and length scales are rescaled. These effective scales are given by
\begin{equation} \label{eq:scale}
\text{Ry}^* = \frac{e^4 m^*}{2 \ka^2 \h^2}, \quad \qquad a^*_B = \frac{\h^2 \ka}{m^* e^2},
\end{equation}
representing the effective Rydberg and Bohr radius, respectively.  

\setlength{\tabcolsep}{5.5pt}
\begin{table}[b]
\begin{center}
\begin{tabular}{l c c c r} 
\hline
\hline
 Semiconductor                  & $\kappa$ & $m^* / m^{}_e$ &  $\text{Ry}^*$ (meV)  &  $a_B^* (\mathrm{A})$ \\
\hline
Cadmium telluride &10.6 & 0.11 & 14.0 & 51.0 \\
Gallium arsenide &13.1 & 0.067 & 5.84 & 103.4\\
Indium phosphide &12.6 & 0.073 & 7.14 & 91.3\\
\hline
\hline
\end{tabular}
\caption{\label{tb:isotropic}Characteristic length and energy scales for impurity states in semiconductors with nondegenerate and isotropic conduction band minima \cite{peter2010fundamentals,haynes2016crc}.}
\end{center}
\end{table}

\begin{figure*}[t]
\begin{center}
\includegraphics[width=0.825\linewidth]{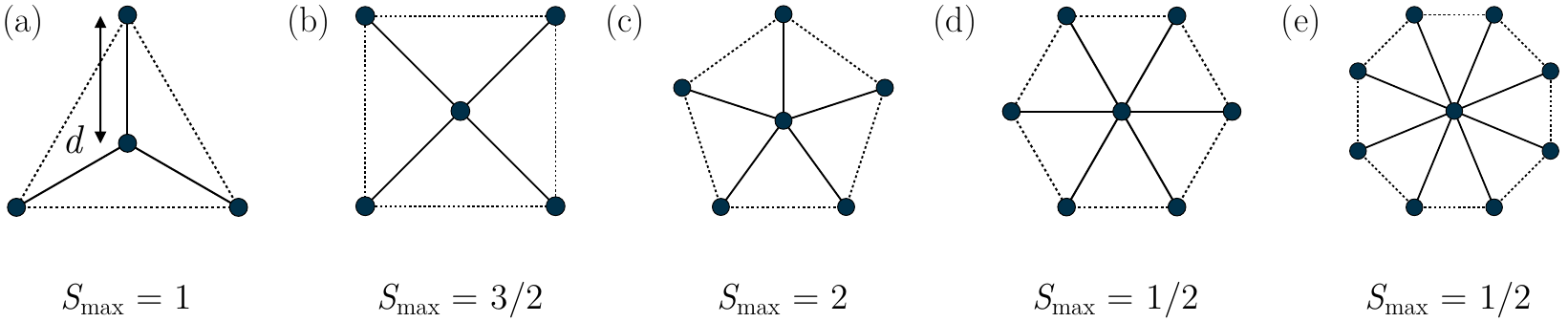}
\caption{Wheel-shaped clusters with (a) three, (b) four, (c) five, (d) six, and (e) eight sites on the rim. When the hopping amplitude between the edge sites is sufficiently small, the central electron forms singlets with the ones along the perimeter. As a result, the edge spins align with each other, yielding a net high-spin magnetic state. For each cluster, we note here the maximum possible value of the ground-state spin, $S_{\mathrm{max}}$.}
\label{fig:wheel} 
\end{center}
\end{figure*}

As an example, for gallium arsenide (GaAs), $\ka = 12.6$ and $m^* = 0.066\, m_e$, where $m_e$ is the free electron mass. The corresponding effective Bohr radius is $a_B^* \sim 100\,$A, which is much larger than the lattice constant $a_L = 5.7\,$A. Thus, the volume occupied by the impurity wavefunction exceeds the unit cell volume containing the dopant by several orders of magnitude. This \textit{a posteriori} justifies neglecting the short-range details of the impurity potential in Eq.~\eqref{eq:qualEq}. Moreover, the energy scale of the impurity states, set by $\text{Ry}^*$, is far smaller than the semiconductor band gap. In this sense, the host lattice effectively serves as an inert vacuum for the impurity problem \cite{Kohn}.  

Such a hydrogenic description of shallow impurities can also be more formally derived within the effective mass approximation, as detailed in Refs.~\onlinecite{Kohn,Shklovskii}. This applies to many semiconductors, such as those noted in Table~\ref{tb:isotropic}, each of which possesses a single conduction band minimum located at the $\Gamma$ point of the Brillouin zone.

As a representative example, let us consider GaAs to illustrate the general mechanism underlying the formation of high-spin clusters. Gallium arsenide is a prototypical zincblende semiconductor, with its lattice structure shown in Fig.~\ref{fig:directLattice}(a). Since GaAs has only one nondegenerate, isotropic valley at the Brillouin zone center, the associated constant-energy surfaces of the conduction band are spherical and centered at $\mathbf{k}\!=\!0$ (cf.~Fig.~\ref{fig:reciprocalLattice} below). Although our primary interest is in many-valley semiconductors with multiple degenerate and equivalent conduction band minima (and anisotropic valleys), we take GaAs here as a benchmark system to demonstrate how high-spin magnetic ground states can arise.

To begin, we study a family of two-dimensional ``wheel-shaped'' clusters sketched in Fig.~\ref{fig:wheel}. Each cluster consists of a regular polygon of sites (the ``rim'') together with one additional site at the center (the ``hub''). The bonds connecting the hub to the rim are referred to as radial connections or ``spokes'' hereafter. A qualitative argument now shows that such charge-neutral  clusters can produce high-spin ground states under suitable conditions. In the uncompensated ground state, each site hosts exactly one electron to minimize the onsite interaction energy $U$. If we set the hopping amplitude between rim sites to zero, by hand, the cluster reduces to a collection of hydrogen molecules sharing a common central site. Recognizing that the hydrogen molecule has a singlet ground state, we see that each rim electron pairs antiferromagnetically with the hub electron. Consequently, the rim spins all align with one another, leading to a collective high-spin ground state for the entire cluster.

When the hopping amplitude along the rim is restored, the ground state \textit{may} instead favor a low-spin configuration. We therefore expect a transition from high spin to low spin as the ratio of rim hopping to radial hopping increases. The argument above holds as long as this ratio is small.  

Physically, geometric considerations constrain the possible high-spin clusters. For rims with five sites or fewer, the nearest-neighbor distance along the rim exceeds the radial bond length. Since hopping amplitudes decay exponentially with distance, hopping between adjacent edge sites is (relatively) suppressed, and high-spin ground states are favored. Once the outer edge comprises six or more sites, however, the radial bond is always longer than or equal to the edge bond length, making the rim hopping significant regardless of cluster size. Hence, neutral wheel-shaped clusters can only realize high-spin ground states when the number of rim sites is strictly less than six.  

This intuition was quantitatively verified by \citet{zhou2013magnetic} using calculations on a generalized Hubbard model \cite{nielsen2007nanoscale,nielsen2010search}. These numerics reveal a clear trend: 
\begin{itemize}
\item 
The wheel-shaped cluster with three edge sites always exhibits a high-spin ground state, independent of size (which is set by the radial length $d$; see Fig.~\ref{fig:wheel}). 
\item 
The cluster with four rim sites supports a high-spin ground state only when $d$ is sufficiently large and above a certain threshold. 
\item
With five sites on the rim, the cluster has a singlet ground state for small sizes, but undergoes successive transitions to a triplet and eventually to a spin-$2$ state as the cluster is enlarged. 
\item
In stark contrast to the previous three cases, the six-edge-site cluster consistently exhibits a low-spin ground state for all $d$. 
\end{itemize}
Furthermore, the value of $d$ at which the transition between the low-spin and high-spin ground states occurs increases with the number of rim sites as the separation between neighboring sites on the edge approaches the radial bond length.

\begin{figure}[t]
\begin{center}
\includegraphics[width=0.9\linewidth]{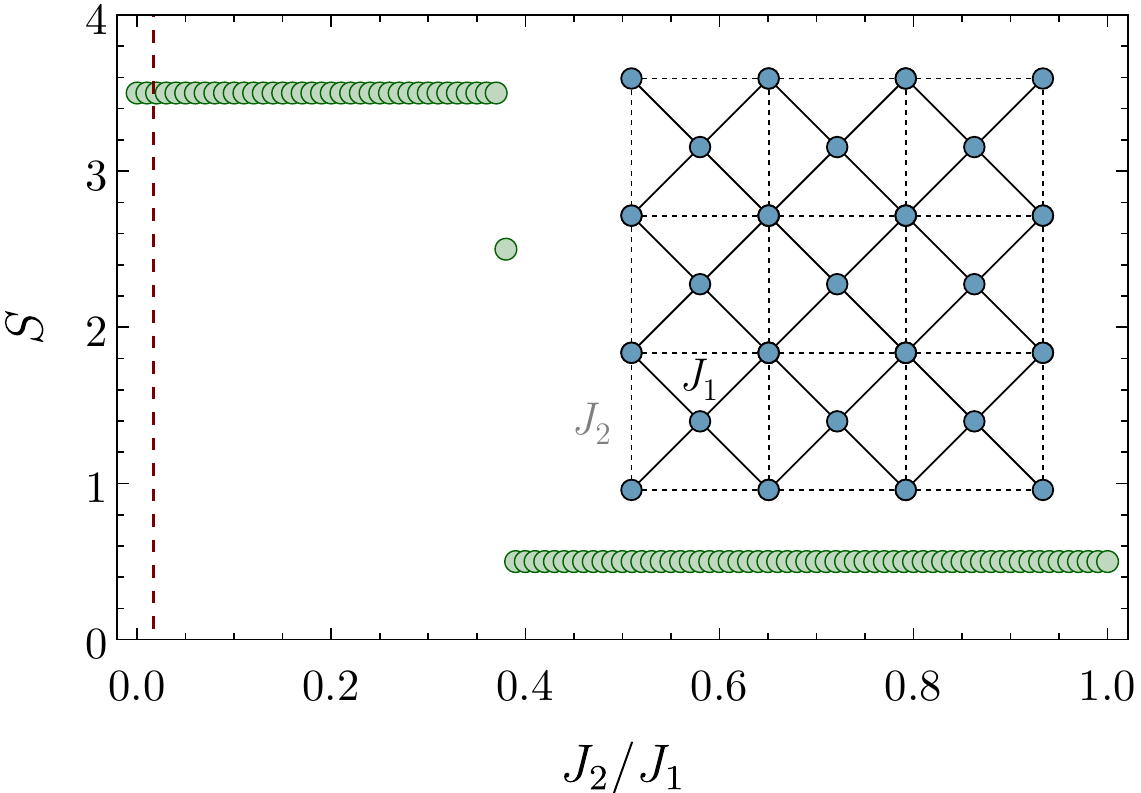}
\caption{Total spin of the ground state of the $J_1$--$J_2$ Heisenberg model on a 25-site cluster as a function of $J_2/J_1$, obtained via exact diagonalization. The inset depicts the lattice structure for nine plaquettes, with $J_1$ and $J_2$ bonds shown as solid and dashed lines, respectively. The red vertical dashed line indicates the value of $J_2/J_1$ realized in the square-lattice geometry when nearest-neighbor sites are placed $6\, a^{*}_B$ apart.}
\label{fig:GaAs} 
\end{center}
\end{figure}

Individual wheel-shaped clusters can be further assembled into larger structures that stabilize even higher total spins. To exemplify this process, we focus on the limit where the onsite interaction $U$ (which energetically disfavors double occupancy of a site) is much larger than the hopping scale $t$. In practice, this can be achieved by placing the dopants sufficiently far apart, typically on the order of $5$--$6\,a^{*}_B$; as $t$ decays exponentially with intersite distance, such a configuration will lie deep in the localized regime. When $U\!\gg\!t$ in a system at charge neutrality, the Hubbard model reduces to a spin-$1/2$ Heisenberg model with an antiferromagnetic exchange interaction $J \sim t^2/U$. Since the electrons are not itinerant, the only interactions between them are magnetic.

The precise details of the resultant magnetic Hamiltonian are lattice-dependent.
For concreteness, let us take a decorated square lattice formed by tiling five-site wheels, as shown in the inset of Fig.~\ref{fig:GaAs}. As the hub-to-rim distance differs from the distance between nearest-neighboring rim sites, the system consists of two \textit{inequivalent} antiferromagnetic bonds, with their relative strengths governed by the geometry. These competing interactions are naturally described by a $J_1$--$J_2$ Heisenberg model,
\begin{equation}
\label{eq:J1J2}
H = J^{}_1 \sum_{\langle i,j \rangle} \mathbf{S}^{}_i \cdot \mathbf{S}^{}_j 
  + J^{}_2 \sum_{\langle\langle i,j \rangle\rangle} \mathbf{S}^{}_i \cdot \mathbf{S}^{}_j,
\end{equation}
where the first and second sums run over nearest- and next-nearest-neighbor bonds, respectively. 
Using exact diagonalization on a 25-site cluster (Fig.~\ref{fig:GaAs}), we find that over a broad range of $J_2/J_1$ values, the ground state stabilizes a sizable total spin of $S=7/2$. 
The asymptotic form of the exchange for large distances $r$ is given by \citet{Herring} as
\begin{equation}
\label{eq:ratio}
J(r) \simeq  1.636 \left(\frac{r}{a^*_B}\right)^{5/2} e^{-2r/a^*_B}
\end{equation}
in Rydbergs. Using this form, for a nearest-neighbor separation of $6\, a^{*}_B$, we find the ratio $J_2/J_1$ 
to be $1.65\times10^{-2}$, which places the system well within the high-spin phase. The same is true using the numerically determined $J(r)$ \cite{kol1965potential} as well. While this simple geometry, with nine square plaquettes glued together, is useful as a toy example, more efficient strategies for assembling large high-spin clusters from wheel-shaped building blocks will be discussed later in the context of Fig.~\ref{fig:square}.

\section{Indirect bandgap semiconductors}
\label{sec:math}

In many common semiconductors, such as silicon and germanium, the conduction band contains several degenerate minima (valleys) at symmetry-related points in the Brillouin zone. As these valleys are displaced from the Brillouin-zone center, the wavefunctions of impurity states are linear combinations of the wavefunctions derived from each of the (anisotropic) conduction band minima, and consequently, deviate significantly from the simple isotropic hydrogenic form. As a result, impurity clusters in such semiconductors display properties that are absent in the hydrogenic clusters discussed above. In this section, we explore the new possibilities enabled by valley degeneracy, particularly in the context of engineering high-spin clusters. Based on the lessons learnt from this exercise, we will then present, in Sec.~\ref{sec:indirect}, examples of high-spin cluster geometries that cannot be realized in semiconductors with a single, nondegenerate conduction band minimum.  

\subsection{Degenerate conduction band minima}

Let $\Psi_j(\r)$ denote the impurity wavefunction associated with the $j$-th conduction band minimum. We orient the coordinate system so that this particular valley is centered at $\mathbf{k}_j = (0,0,k_0)$ in the Brillouin zone. When the impurity wavefunction extends over a region much larger than a lattice cell, the effective mass approximation allows us to write it as \cite{Kohn}
\begin{equation}
\Psi^{}_j(\r) = F^{}_j(\r)\, u^{}_j(\r)\, e^{i \mathbf{k}_j \cdot \r},
\end{equation}
where $u_j(\r)$ is the periodic Bloch function at $\mathbf{k}_j$.  

Since $\mathbf{k}_j \neq 0$, the dispersion near the valley minimum is anisotropic:  
\begin{equation}
E(\mathbf{k}) = \frac{\hbar^2}{2m^{}_l}(k^{}_z - k^{}_0)^2 + \frac{\hbar^2}{2m^{}_t}(k_x^2 + k_y^2),
\end{equation}
with $m_l$ and $m_t$ denoting the longitudinal and transverse effective masses. The corresponding envelope function satisfies the anisotropic effective mass equation
\begin{equation} \label{eq:anisoH}
\left[-\frac{\hbar^2}{2m^{}_l} \pd{^2}{z^2} - \frac{\hbar^2}{2m^{}_t} \left(\pd{^2}{x^2} + \pd{^2}{y^2}\right) - \frac{e^2}{\kappa r} - E \right] F^{}_j(\r) = 0,
\end{equation}
where $\kappa$ is the dielectric constant of the host. As listed in Table~\ref{table:prop}, we often find $m_l \gg m_t$ in typical semiconductors, so the impurity wavefunction extends much farther in the transverse plane than along the longitudinal axis. Intuitively, this follows from the Bohr radius scaling as $a^{*}_B \propto 1/m^\ast$ in the isotropic hydrogenic limit.  

The ground-state envelope $F_j(\r)$ can be accurately obtained using the variational principle with the trial function \cite{Faulkner}
\begin{equation} \label{eq:anisoF}
F^{}_j(\r) = \left(\frac{1}{\pi a^2 b}\right)^{1/2} 
\exp\!\left[-\sqrt{\frac{z^2}{b^2} + \frac{x^2 + y^2}{a^2}}\right],
\end{equation}
where $a$ and $b$ are variational parameters. In the isotropic case ($m_t = m_l$), the variational calculation of course reproduces the exact hydrogenic result with $a = b = a^{*}_B$.  

\begin{figure*}[t]
\begin{center}
\includegraphics[width=\linewidth]{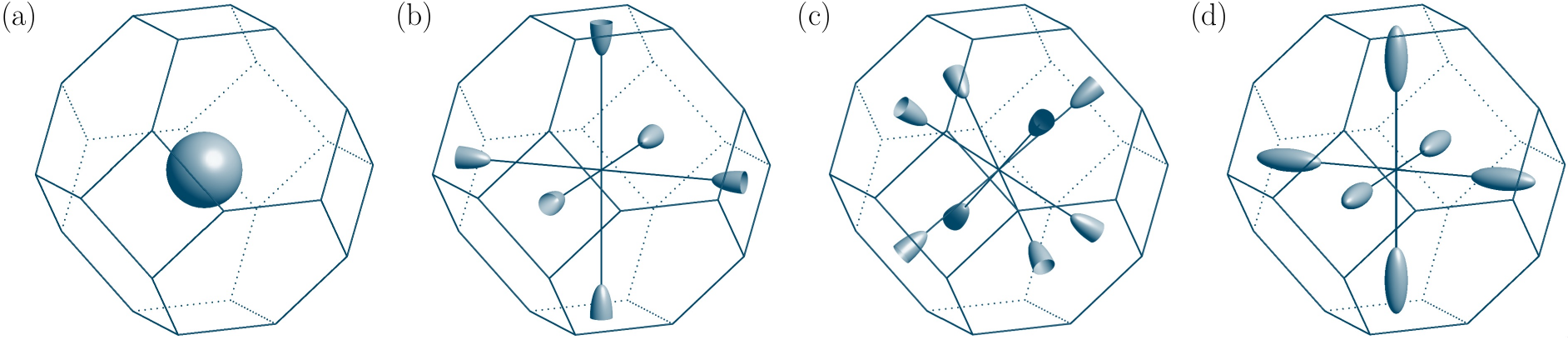}
\caption{Constant-energy surfaces in the Brillouin zone of (a) GaAs (b) AlAs, (c) Ge, and (d) Si, exhibiting one, three, four, and six valleys, respectively. The conduction band minima are located at the centers of the ellipsoids.}
\label{fig:reciprocalLattice}
\end{center}
\end{figure*}

\setlength{\tabcolsep}{8.8pt}
\begin{table}[tb]
\centering
\begin{tabular}{ l c c c } 
\hline
\hline
Semiconductor  & $m^{}_l$ & $m^{}_t$ & $m^{}_t/m^{}_l$ \\
\hline
Aluminum arsenide & $1.050 \, m^{}_e$ & $0.205 \, m^{}_e$ & 0.195 \\
Germanium         & $1.588 \, m^{}_e$ & $0.082 \, m^{}_e$ & 0.051 \\
Silicon           & $0.916 \, m^{}_e$ & $0.191 \, m^{}_e$  & 0.208 \\
\hline
\hline
\end{tabular}
\caption{\label{table:prop}Effective mass parameters for representative semiconductors with degenerate anisotropic conduction band minima \cite{Faulkner, Padmanabhan}.}
\end{table}

If the conduction band minimum is $\mathcal{N}$-fold degenerate, the effective mass Hamiltonian admits an $\mathcal{N}$-fold degenerate lowest-energy manifold. In practice, this degeneracy is lifted by the short-range central-cell potential of the impurity atom, which is not captured within the effective mass approximation. The true ground state is a linear combination of the valley wavefunctions,  
\begin{equation}
\Psi(\r) = \sum_{j=1}^\mathcal{N} \alpha^{}_j \Psi^{}_j(\r) 
= \sum_{j=1}^\mathcal{N} \alpha^{}_j F^{}_j(\r)\, u^{}_j(\r)\, e^{i \mathbf{k}_j \cdot \r},
\end{equation}
where the coefficients $\alpha_j$ are, in general, complex amplitudes \cite{Kohn}. In most cases, the central-cell potential is attractive, and the ground state is a symmetric superposition with equal weight from all equivalent conduction band minima.

\subsection{Hopping integrals}

Having written down the impurity wavefunctions, we can now calculate the hopping parameter for anisotropic clusters using a tight-binding model. The tight-binding approximation is valid when electrons in the cluster remain localized on individual impurity sites and interact only weakly with neighboring sites. In this regime, the electronic energies are close to those of isolated atomic orbitals, and the overlap between wavefunctions centered on different sites is small. The dominant effect of a neighboring site at a displacement $\mathbf{R}$ is then to introduce a mixing matrix element
\begin{equation}
t(\mathbf{R}) = \bra{\Psi(\r)} \Delta \mathcal{V}(\r) \ket{\Psi(\mathbf{r} - \mathbf{R})},
\end{equation}
where $\Delta \mathcal{V}$ represents the additional potential contributed by the atom situated at $(\mathbf{r} - \mathbf{R})$ \cite{Mott}. This quantity plays the same role as the hopping parameter in the Hubbard model and we hereafter refer to it as the hopping integral.  

For the case of two atoms in a hydrogenic molecule, the hopping integral decays exponentially with separation and is given by
\begin{equation}
t(R) = 2 \left(1 + \frac{R}{a^{*}_B}\right) e^{-R/a^{*}_B},
\end{equation}
in units of Rydbergs \cite{Slater}.  
In our case however, the electronic wavefunction contains contributions from multiple conduction band valleys, and the hopping integral takes the more general form
\begin{equation} \label{eq:hop}
t(\mathbf{R}) = \sum_{j=1}^\mathcal{N} \sum_{l=1}^\mathcal{N} \alpha_j^\ast \alpha^{}_l 
\bra{\Psi^{}_j(\r)} \frac{1}{|\mathbf{r} - \mathbf{R}|} \ket{\Psi^{}_l(\mathbf{r} - \mathbf{R})},
\end{equation}
where we have omitted the constant prefactor $e^2 / \kappa$ for notational convenience.  

Unlike in the hydrogenic case, $t(\mathbf{R})$ is no longer solely a function of $|\mathbf{R}|$, since the impurity wavefunctions are anisotropic. Moreover, each valley wavefunction $\Psi_j(\r)$ carries a complex phase factor $\exp({i \mathbf{k}_j \cdot \r})$, which allows for constructive or destructive interference between different terms in the double sum. As a result, the hopping parameter can be heavily suppressed for suitable choices of impurity placement. This feature is particularly useful for engineering magnetic clusters (as we show in Sec.~\ref{sec:indirect}), since eliminating certain electron hoppings enhances the likelihood of achieving a high-spin ground state.

\subsection{Candidate materials}
In the next section, we develop methods to exploit interference effects in the sum for $t(\mathbf{R})$, allowing us to suppress the hopping parameter between selected pairs of impurity sites. This construction depends sensitively on the underlying lattice and band structures of the host semiconductor. Therefore, before proceeding further, we provide a brief review of these properties for several representative materials.

Figure \ref{fig:directLattice} shows the unit cells of crystals with diamond cubic and zincblende structures. The diamond cubic lattice consists of two interpenetrating face-centered cubic (FCC) sublattices displaced along the body diagonal of the cubic cell by one quarter of its length. The zincblende structure has the same geometric arrangement, but the two interpenetrating sublattices consist of atoms of different species. In both cases, each atom has four nearest neighbors arranged at the vertices of a regular tetrahedron; for zincblende, these neighboring atoms are of distinct elements. 

In reciprocal space, the Brillouin zone of semiconductors with diamond cubic or zincblende direct lattices has the shape of a truncated octahedron \cite{Ashcroft}. Figure \ref{fig:reciprocalLattice} illustrates the constant-energy surfaces in the Brillouin zone for several semiconductors with these crystal structures. Specifically, we consider aluminum arsenide (AlAs), germanium (Ge), and silicon (Si), which have three, four, and six valleys, respectively.

Silicon and germanium are among the most extensively studied semiconductors, both theoretically and experimentally. Both have a diamond cubic structure and associatedly, truncated octahedral Brillouin zones \cite{Madelung}. Silicon hosts six conduction band minima located inside the Brillouin zone along the cubic axes, i.e., the $(\pm1,0,0)$, $(0,\pm1,0)$, and $(0,0,\pm1)$ directions. In contrast, germanium has four conduction band minima situated on the Brillouin zone boundary along the cubic diagonals, i.e., $(1,1,1)$, $(1,-1,-1)$, $(-1,1,-1)$, and $(-1,-1,1)$. In both cases, the valleys are anisotropic, as they do not lie at the zone center.

Aluminum arsenide, though less commonly studied, exhibits a band structure closely resembling that of silicon. Its direct lattice has zincblende symmetry, with a truncated octahedral Brillouin zone. The conduction band minima are located along the cubic axes, but unlike silicon, they lie at the Brillouin zone boundary. Consequently, there are only three distinct minima, since opposite zone boundaries are equivalent.\\

\section{Clusters in multivalley semiconductors}
\label{sec:indirect}

With this background, we now construct explicit examples to demonstrate how interference can be used to suppress hopping between pairs of impurity sites. The key idea is that if the hopping between certain sites is canceled out, so is the antiferromagnetic spin exchange between them. We begin with the relatively simple case of aluminum arsenide and then turn to the more intricate examples of germanium and silicon.

\vspace*{0.25cm}
\subsection{Aluminum arsenide}  

Aluminum arsenide has three conduction band minima in the Brillouin zone, located at  
\begin{equation}
\mathbf{k}^{}_1 = \frac{2\pi}{a^{}_L} (1, 0, 0), \,  
\mathbf{k}^{}_2 = \frac{2\pi}{a^{}_L} (0, 1, 0), \,
\mathbf{k}^{}_3 = \frac{2\pi}{a^{}_L} (0, 0, 1),   
\end{equation}
where $a^{}_L$ denotes the lattice constant. The ground-state impurity wavefunction is given by  
\begin{equation}
\label{eq:psiAlAs}
\Psi(\r) = \sqrt{\frac{1}{3}} \sum_{j = 1}^3 F^{}_j(\r)\, u^{}_j(\r)\, e^{i \mathbf{k}_j \cdot \r}.
\end{equation}

To ensure equivalence among impurity centers, we consider placing dopants only on the sites of a single atomic species in the zincblende lattice. For example, within the unit cell that has an aluminum atom at the origin, donor impurities such as silicon may be positioned at aluminum sites located at  
\begin{equation*}
a^{}_L \left(0, 0, 0\right), \, \,
a^{}_L \left(0, \frac{1}{2}, \frac{1}{2} \right), \, \,
a^{}_L \left(\frac{1}{2}, 0, \frac{1}{2} \right), \,  \,
a^{}_L \left(\frac{1}{2}, \frac{1}{2}, 0 \right).
\end{equation*}

Inserting Eq.~\eqref{eq:psiAlAs} into \eqref{eq:hop}, the hopping integral reads
\begin{widetext}
\begin{alignat}{1} 
t(\mathbf{R}) = \frac{1}{3} \sum_{j= 1}^3 \sum_{l = 1}^3 &\int d^3\r\, e^{-i(\mathbf{k}_j - \mathbf{k}_l) \cdot \r} e^{i \mathbf{k}_l \cdot \mathbf{R}} u^*_j(\r)\, u^{}_l(\mathbf{r} - \mathbf{R})\, F^*_j(\r)\, F^{}_l(\mathbf{r} - \mathbf{R})\, \frac{1}{|\mathbf{r} - \mathbf{R}|}.
\end{alignat}
For $j$\,$\neq$\,$l$, the exponential factor $\exp[-i(\mathbf{k}_j - \mathbf{k}_l) \cdot \r]$ oscillates rapidly, causing the integral to effectively vanish. Thus, the expression above simplifies to  
\begin{alignat}{1} 
t(\mathbf{R}) = \frac{1}{3} \sum_{j = 1}^3 e^{i \mathbf{k}_j \cdot \mathbf{R}} &\int d^3\r\, u^*_j(\r)\, u^{}_j(\mathbf{r} - \mathbf{R}) \, F^{}_j(\r)\, F^{}_j(\mathbf{r} - \mathbf{R})\, \frac{1}{|\mathbf{r} - \mathbf{R}|}. 
\end{alignat}
Crucially, the phase factor outside the integral can induce partial cancellation among the terms in this sum.\\

Note that if the wavefunctions $u_j(\r) F_j(\r)$ were isotropic, the integral itself would take the same value for all $j$. The anisotropy, however, complicates matters as the integrals now depend explicitly on the orientation of $\mathbf{R}$, requiring a separate treatment for each $j$.  The Bloch functions $u_j$ are periodic on the lattice, and the envelope functions $F_j$ vary on the much longer length scale $a^*_B$, i.e., they are essentially constant inside the unit cell. Hence, the integral \textit{within} a unit cell can be done first, and then the remaining sum \textit{over} unit cells can be replaced by an integral. \\

Carrying out this procedure, without loss of generality, let us single out the long-distance contribution for the valley along $(1,0,0)$. To do so, we transform to a new set of variables $\mathbf{R}' = (X/b, Y/a, Z/a)$ and $\r' = (x/b, y/a, z/a)$, in terms of which, we define  
\begin{alignat}{1} 
\xi^{}_1(\mathbf{R}) &\equiv \int d^3\r\, F^{}_1(\r)\, F^{}_1(\r- \mathbf{R})\, \frac{1}{|\mathbf{r} - \mathbf{R}|}= \frac{1}{a^2 b} \int d^3\r'\, e^{-|\r' - \mathbf{R}'|} e^{-r'}\, V^{}_1(\r' - \mathbf{R}').
\label{eq:s(R)}
\end{alignat}
The anisotropy resides entirely in the interaction term, 
\begin{equation} 
 V^{}_1(\r' - \mathbf{R}') \equiv\left[ \frac{1}{b^2(x' - X')^2 + a^2(y' - Y')^2 + a^2(z' - Z')^2} \right]^{1/2}.
\end{equation}
As a first approximation, we temporarily neglect this anisotropy by writing $V_1(\r' - \mathbf{R}') = V(|\r' - \mathbf{R}'|)$ with $V(r) = 1/(ar)$. This allows the expressions for all three valleys to be written in a common form, 
\begin{align}
\xi^{}_1(\mathbf{R}) = \xi \!\left( \sqrt{\frac{X^2}{b^2} + \frac{Y^2}{a^2} + \frac{Z^2}{a^2}} \,\right), \quad
\xi^{}_2(\mathbf{R}) = \xi \!\left( \sqrt{\frac{X^2}{a^2} + \frac{Y^2}{b^2} + \frac{Z^2}{a^2}} \,\right), \quad 
\xi^{}_3(\mathbf{R}) = \xi \!\left( \sqrt{\frac{X^2}{a^2} + \frac{Y^2}{a^2} + \frac{Z^2}{b^2}} \,\right),  
 \label{eq:snum}
\end{align}
where  
\begin{equation}
\xi(R) \equiv \frac{1}{a^2 b} \int d^3\r\, e^{-|\mathbf{r} - R\, \hat{\boldsymbol{n}}|} e^{-r}\, V(|\mathbf{r} - R\, \hat{\boldsymbol{n}}|),
\end{equation}
and $\hat{\boldsymbol{n}}$ is an arbitrary unit vector.
\end{widetext}  

\begin{figure*}[tb]
    \centering
    \includegraphics[width=0.8\linewidth]{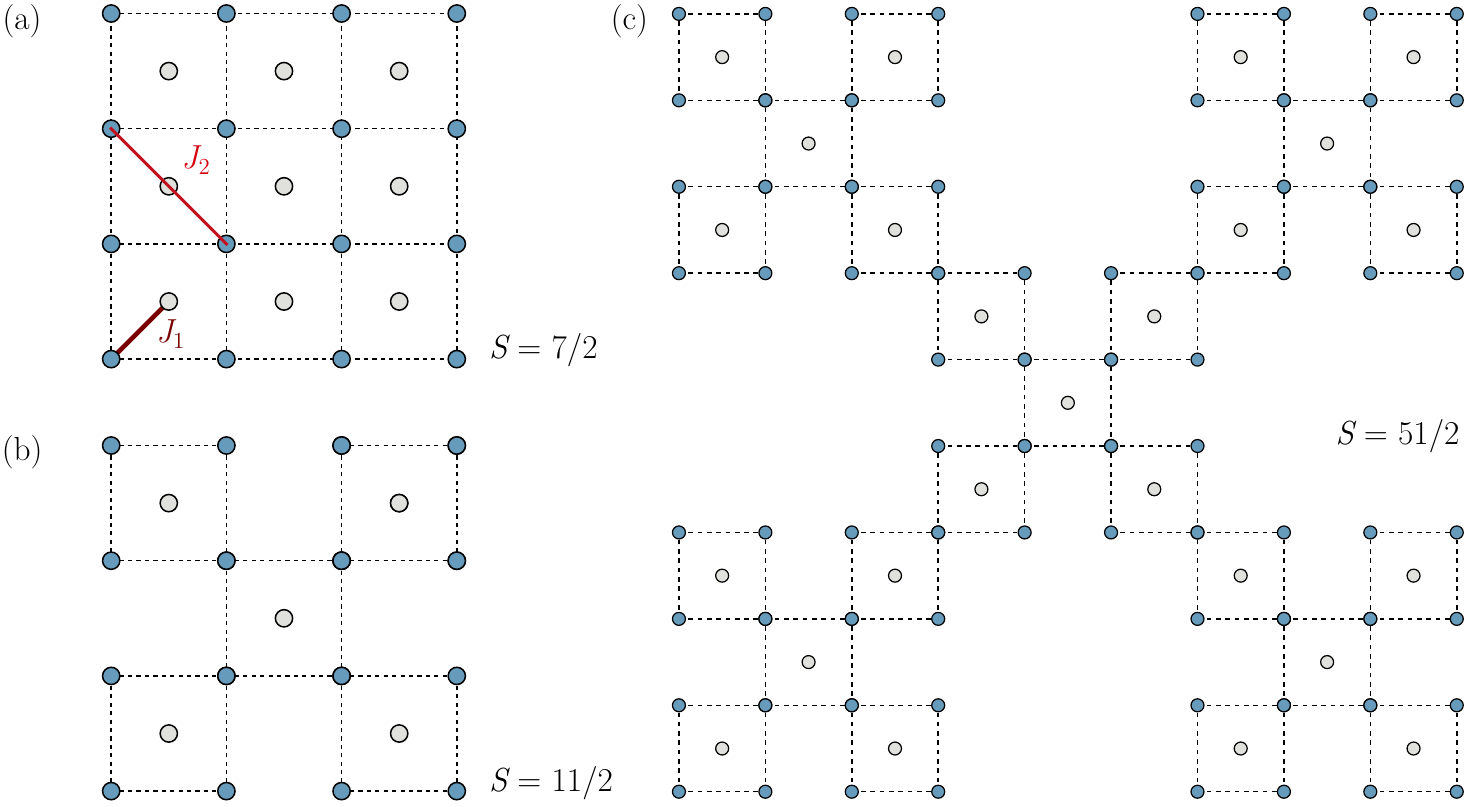}
    \caption{(a) Decorated square lattice obtained by tiling the two-dimensional plane with five-site wheel-shaped clusters. By virtue of cancellation of the hopping integrals, the exchange interaction is nearly suppressed along the vertical and horizontal bonds (dashed lines). Representative nearest- and next-nearest-neighbor spin exchanges are labeled $J_1$ and $J_2$, respectively, and indicated by solid red lines. Unlike in Fig.~\ref{fig:GaAs}, here, the next-nearest-neighbor bonds are twice the length of the nearest-neighbor ones (rather than $\times \sqrt{2}$ previously). (b) Generation $g=1$ cluster formed by arranging square plaquettes in a fractal-like pattern. Despite having fewer sites (21) than the uniform lattice in (a) (25), this cluster has a larger total spin of $S=11/2$ compared to $S=7/2$, owing to its greater perimeter contribution. Note that this cluster could also be tiled uniformly by translations to form an extended checkerboard lattice, rather than iterated fractally. (c) Generation $g=2$ fractal-like cluster, obtained by repeating the tiling process with the $g=1$ structure as the elemental unit. With $101$ sites, we attain a net ground-state spin of $S=51/2$. All of these high-spin configurations---whether realized as extended lattices or as fractal structures---can also be implemented with GaAs as the host semiconductor.}
    \label{fig:square}
\end{figure*}

Putting everything together, the hopping integral can be expressed as  
\begin{equation} \label{eq:t(R)}
t(\mathbf{R}) = \frac{1}{3} \sum_{j = 1}^3 e^{i \mathbf{k}_j \cdot \mathbf{R}}\, \xi^{}_j(\mathbf{R}).
\end{equation}
This form closely resembles the hopping integral for impurity clusters in silicon derived in Ref.~\onlinecite{Bhatt2}. A related expression for the exchange interaction $J(\mathbf{R})$ can be found in Appendix~A of Ref.~\onlinecite{Andres}, which displays the same interference factor as obtained from Eq.~\eqref{eq:t(R)}. A comparison of various approximations for the exchange \cite{angelescu2002effective} shows them all decaying exponentially with distance with the \textit{same} exponent.

The function $\xi(R)$ decreases extremely rapidly with $R$ because the overlap factor $e^{-|r - R|} e^{-r}$ decays exponentially. For example, if the separation vector $\mathbf{R}$ lies along the $x$-axis, Eq.~\eqref{eq:snum} implies that $\xi_1 \ll \xi_2 = \xi_3$, since $a > b$.  This property has important implications.
Consider, for instance, two sites separated by the vector  
\begin{equation}
\mathbf{R} = a^{}_L (n^{}_x, 0, 0) + a^{}_L \left(\frac{1}{2}, \frac{1}{2}, 0 \right),
\end{equation}
where $|n_x|$ is an integer much larger than one, so that $\mathbf{R}$ is essentially along $\hat{x}$. From Eq.~\eqref{eq:t(R)}, it follows that  
\begin{equation}
e^{i \mathbf{k}_2 \cdot \mathbf{R}} \xi^{}_2(\mathbf{R}) = -\, e^{i \mathbf{k}_3 \cdot \mathbf{R}} \xi^{}_3(\mathbf{R}),
\end{equation}
so that the $j = 2,3$ terms cancel. Consequently, $\lvert t(\mathbf{R})\rvert$ is determined primarily by $\xi_1(\mathbf{R})$, which is much smaller than the total magnitude of the terms in the sum \eqref{eq:t(R)}. Hence, electron hopping along the $x$-axis can be exponentially suppressed by a suitable choice of impurity placement within the semiconductor lattice.  
This result remains valid even when we account for the anisotropic contributions from the  factors of $V_j(\mathbf{r} - \mathbf{R})$ that were previously neglected. Our earlier approximation introduces the same error in $\xi_j(\mathbf{R})$ for all directions within the transverse plane, so we still have $\xi_2 = \xi_3$ for hopping parallel to the $x$-axis, ensuring cancellation of the $j=2,3$ terms in Eq.~\eqref{eq:t(R)}. Moreover, $\xi_1$ remains much smaller than $\xi_2$ and $\xi_3$, since the dominant exponential terms in the integral defining $\xi_j(\mathbf{R})$ are unaffected by the approximation.

Similarly, for two sites separated primarily along the $y$-axis, we obtain $\xi_2 \ll \xi_1 = \xi_3$. The hopping can thus be nearly zeroed out by positioning two impurities with separation  
\begin{equation}
\mathbf{R} = a^{}_L (0, n^{}_y, 0) + a^{}_L \left(\frac{1}{2}, \frac{1}{2}, 0 \right),
\end{equation}
where $|n_y| \gg 1$, which facilitates the cancellation of the $j = 1, 3$ terms. This allows the construction of impurity clusters in the two-dimensional $x$--$y$ plane, with strategically placed dopants to suppress hoppings almost parallel to either coordinate axis.

As a concrete example, consider the wheel-shaped cluster shown in Fig.~\ref{fig:wheel}(b). 
The ``rim'' of the wheel is formed by dopants placed, say, at 
$(0,0,0)$, $a^{}_L(n+1/2,\,1/2,\,0)$, $a^{}_L(1/2,\,n+1/2,\,0)$, and $a^{}_L(n,\,n,\,0)$, 
while the ``hub'' is positioned at $a^{}_L(n/2,\,(n+1)/2,\,1/2)$, with $n \in 2 \mathbb{Z}$. 
In practice, we choose $n$ such that $a^{}_L n \approx 6\,a^{*}_B$. 
If the dopants are placed much farther apart than this characteristic scale, the desirable exchange interaction, which also decays exponentially with distance, becomes negligibly small (compared to the temperature). 
Conversely, if they are placed too close together, the electrons delocalize and the system crosses over to an itinerant regime. 
As argued above, the hoppings between adjacent sites on the rim of the wheel are exponentially suppressed. For the radial hoppings, one can easily check that $\xi_1 = \xi_2 \ll \xi_3$, so the $j\!=\!1,2$ terms again cancel. Therefore, the resulting magnitude of these hoppings is $\xi_3$, which is not significantly
different from the sum of the magnitudes of $\xi_1$, $\xi_2$, and $\xi_3$. In the purely hydrogenic case, this cluster exhibits a ground state with $S=1/2$ at small sizes, owing to the nonvanishing hopping amplitudes between edge sites \cite{zhou2013magnetic}. Once these hoppings are suppressed, the ground state corresponds to an $S=3/2$ phase.

Thus, our central observation is that since hoppings in \textit{both} the vertical and horizontal directions---and not just along single lines---can be suppressed, we can construct a five-site wheel with a high-spin ground state. This wheel can then serve as a fundamental building block for larger high-spin structures. 
One natural approach is to tile the plane with these wheels. The simplest realization thereof is a decorated square lattice, which contains two sites per unit cell. In this geometry, the spin at the center of each plaquette couples to the four surrounding sites and favors antialigning their spins with its own. As a result, each unit cell effectively carries spin zero. Accordingly, the total spin of a finite cluster is determined only by contributions from the boundary, while the bulk does not contribute. For example, the cluster shown in Fig.~\ref{fig:square}(a) contains 25 dopants but has a total spin of only $S=7/2$. More generally, for a system of $N$ sites, the perimeter contribution scales as $\sqrt{N}$, implying that the spin per site, $s \equiv S/N \sim 1/\sqrt{N}$, vanishes in the thermodynamic limit $N \rightarrow \infty$.

Since the dominant contribution to the net spin arises from the perimeter, a more efficient strategy is to deliberately maximize such surface contributions. This can be achieved by constructing fractal-like clusters. The procedure begins with a single five-site block of the decorated square lattice, to which we assign the  ``generation'' index $g\!=\!0$. To obtain the next generation, identical blocks are attached to the outermost corner sites of the existing cluster. Repeating this process iteratively produces a hierarchy of increasingly larger clusters, where each generation preserves the same basic geometry but extends the boundary. In this way, the fraction of sites that lie on the perimeter remains finite even as the system grows. This iterative construction is illustrated in Fig.~\ref{fig:square}, which shows clusters for generations $g\!=\!1$ (b) and $g\!=\!2$ (c).  For generation $g$, the net spin scales as 
$
S = (2 \cdot 5^{g+1} + 1)/2,
$
while the total number of sites is 
$
N = 4 \cdot 5^{g+1} + 1.
$
Thus, in the limit $g \rightarrow \infty$, the spin per site approaches 
$
s \rightarrow 1/4,
$
corresponding to a magnetization equal to 50\% of that of a fully polarized state. Compared to the uniform decorated square lattice, this construction provides a far more efficient route to building large high-spin clusters.

\subsection{Germanium}  

Germanium has four conduction band minima in the Brillouin zone, located at  
\begin{alignat}{2}
\nonumber
\mathbf{k}^{}_1 &= \frac{\pi}{a^{}_L} (1, 1, 1), \quad & \mathbf{k}^{}_2 &= \frac{\pi}{a^{}_L} (1, -1, -1),  \\
\mathbf{k}^{}_3 &= \frac{\pi}{a^{}_L} (-1, 1, -1), \quad & \mathbf{k}^{}_4 &= \frac{\pi}{a^{}_L} (-1, -1, 1),   
\end{alignat}
so the ground-state impurity wavefunction is  
\begin{equation}
\Psi(\r) = \frac{1}{2} \sum_{j = 1}^4 F^{}_j(\r)\, u^{}_j(\r)\, e^{i \mathbf{k}_j \cdot \r}.
\end{equation}

Since germanium is an elemental semiconductor, donor impurities such as tin or arsenic may be placed at all eight cubic sites within the unit cell. For a cell with one atom situated at the origin, the different sites are located at  
\begin{alignat}{4}
\nonumber
&a^{}_L \,(0, 0, 0), \,
&& a^{}_L \left(0, \frac{1}{2}, \frac{1}{2}\right), \,
&& a^{}_L \left(\frac{1}{2}, 0, \frac{1}{2}\right), \, 
&& a^{}_L \left(\frac{1}{2}, \frac{1}{2}, 0\right), \\
&a^{}_L \left(\frac{1}{4}, \frac{1}{4}, \frac{1}{4}\right), \,
&& a^{}_L \left(\frac{1}{4}, \frac{3}{4}, \frac{3}{4}\right), \, 
&& a^{}_L \left(\frac{3}{4}, \frac{1}{4}, \frac{3}{4}\right), \, 
&& a^{}_L \left(\frac{3}{4}, \frac{3}{4}, \frac{1}{4}\right).
\label{eq:diamond}
\end{alignat}

Following the approach outlined in the previous subsection, we write the hopping parameter as  
\begin{equation} 
\label{eq:t2(R)}
t(\mathbf{R}) = \frac{1}{4} \sum_{j = 1}^4 e^{i \mathbf{k}_j \cdot \mathbf{R}}\, \xi^{}_j (\mathbf{R}),
\end{equation}
where the  matrix element associated with each valley is 
$
\xi_j(\mathbf{R})$\,$=$\,$\int d^3\r F_j(\r) F_j(\mathbf{r} - \mathbf{R})\, {|\mathbf{r} - \mathbf{R}|}^{-1}
$ as derived above.
The functions $F_j(\r)$, which solve the effective mass equation, differ solely in their definition of the longitudinal axis, which is aligned with the direction of $\mathbf{k}_j$. So, $\xi_j(\mathbf{R})$ depends only on the projections of $\mathbf{R}$ along the longitudinal and transverse axes of the $j$-th valley. Moreover, $\xi_j$ is symmetric under reflections of $\mathbf{R}$ about any coordinate axis, so $\xi_j(\mathbf{R}) = \xi_l(\mathbf{R})$ whenever $|\mathbf{k}_j \cdot \mathbf{R}| = |\mathbf{k}_l \cdot \mathbf{R}|$.  \\

\begin{figure*}[tb]
    \centering
    \includegraphics[width=0.95\linewidth]{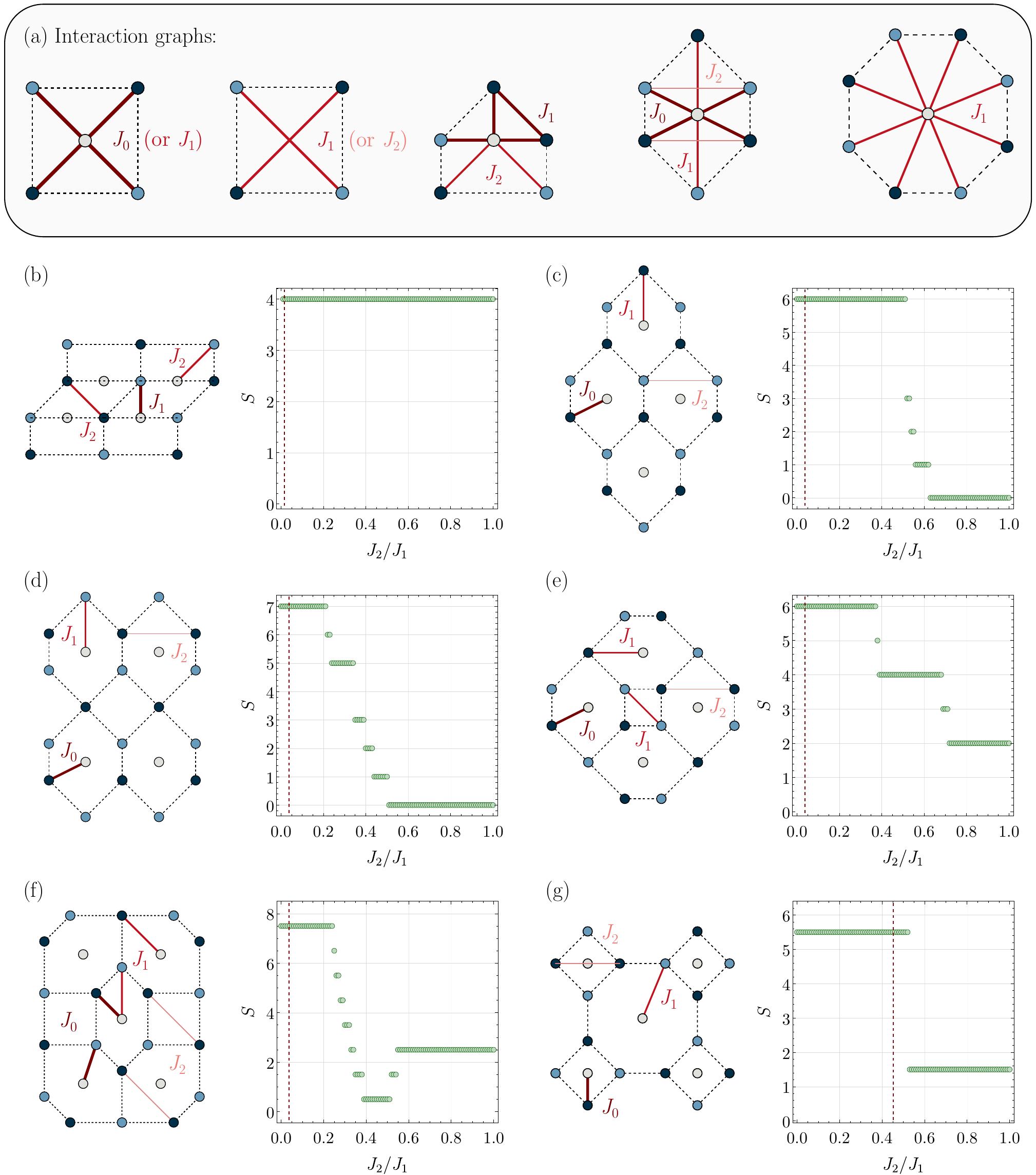}
    \caption{(a) Interaction graphs of wheel-shaped clusters with square, pentagonal, hexagonal, and octagonal geometries. In each unit, the dominant Heisenberg couplings are indicated, with darker colors and thicker lines representing stronger interactions. Light and dark blue circles denote dopant sites with $z$ coordinates equal to $0$ and $1/2$, respectively. Grey circles represent the hub dopants, positioned approximately at the center of each plaquette; their $z$ coordinate can be chosen arbitrarily, subject only to the requirement that the hub--rim interaction does not cancel out.
(b--g) Examples of high-spin dopant clusters constructed from the building blocks in (a). For visual clarity, only one representative interaction is shown for each symmetry-inequivalent bond; the full set of couplings is specified by panel (a). Dashed lines denote couplings that are exponentially suppressed due to cancellation of the hopping integrals. For each cluster geometry, the total spin of the ground state of the $J_1$--$J_2$ Heisenberg model  is computed as a function of $J_2/J_1$ using exact diagonalization. In these calculations, we set $J_0 = J_1$, which underestimates the extent of the high-spin phase. The red vertical dashed line marks the value of $J_2/J_1$ realized in the corresponding geometry when the sites connected by $J_1$ bonds are separated by $6\,a^{*}_B$.  
}
    \label{fig:Ge}
\end{figure*}

We now look at four specific vectors in the $x$--$y$ plane:  
\begin{alignat}{2}
\nonumber
\mathbf{R}^{}_1 &= a^{}_L (1, 0, 0), \quad & \mathbf{R}^{}_2 &= a^{}_L (0, 1, 0), \\
\mathbf{R}^{}_3 &= a^{}_L (1, 1, 0), \quad & \mathbf{R}^{}_4 &= a^{}_L (1, -1, 0).
\label{eq:cancDir}
\end{alignat}
Note that $|\mathbf{R}_1 \cdot \mathbf{k}_j|\!=\!\pi$ for all $j$. Thus, for hopping exactly parallel to the $x$-axis, all the  $\xi_j$ terms in Eq.~\eqref{eq:t2(R)} are equal. Suppose  
\begin{equation}
\label{eq:disp1}
\mathbf{R} = n \mathbf{R}^{}_1 + a^{}_L \left(0, \frac{1}{2}, \frac{1}{2}\right),
\end{equation}
where $\lvert n \rvert \gg 1$; in this case, the sum of the four phase factors $\exp({i \mathbf{k}_j \cdot \mathbf{R}})$ vanishes. Therefore, $t(\mathbf{R}) = 0$ (up to small differences in the $\xi_j(\mathbf{R})$ caused by the slight transverse displacement of $\mathbf{R}$). This shows that electron hopping parallel to the $x$-axis in a germanium lattice can be strongly suppressed. By symmetry, the same conclusion follows for hopping parallel to the $y$-axis using 
\begin{equation}
\label{eq:disp2}
\mathbf{R} = n \mathbf{R}^{}_2 + a^{}_L \left(0,\frac{1}{2},\frac{1}{2}\right).
\end{equation}
Likewise, for hopping along $\mathbf{R}_3$, observe that $|\mathbf{R}_3 \cdot \mathbf{k}_1|\!=\! |\mathbf{R}_3 \cdot \mathbf{k}_4|\!=\! 2\pi$ and $|\mathbf{R}_3 \cdot \mathbf{k}_2|\!=\! |\mathbf{R}_3 \cdot \mathbf{k}_3|\!=\!0$, wherefore $\xi_1 = \xi_4$ and $\xi_2 = \xi_3$. With  
\begin{equation}
\mathbf{R} = n \mathbf{R}_3 + a^{}_L \left(0, \frac{1}{2}, \frac{1}{2}\right), \quad |n| \gg 1,
\end{equation}
we find that $e^{i \mathbf{k}_1 \cdot \mathbf{R}} = -e^{i \mathbf{k}_4 \cdot \mathbf{R}}$ and $e^{i \mathbf{k}_2 \cdot \mathbf{R}} = -e^{i \mathbf{k}_3 \cdot \mathbf{R}}$, so the $j=1,4$ terms and the $j=2,3$ terms cancel pairwise in Eq.~\eqref{eq:t2(R)}. The same reasoning also applies to $\mathbf{R}_4$, thereby accounting for both diagonals $y = \pm x$ in the $x$--$y$ plane.

An important consequence of this cancellation is that in addition to being able to suppress hoppings along the vertical and horizontal directions, now hoppings between sites oriented at $45^\circ$ can also be suppressed. This introduces the possibility of tiling the plane with geometries beyond simple squares, enabling the construction of a wide variety of clusters with distinct high-spin ground states. Examples of such tilings are compiled in Fig.~\ref{fig:Ge}(b--g), which are comprised of the basic plaquettes (squares, pentagons, hexagons, and octagons) shown in Fig.~\ref{fig:Ge}(a). The spin exchanges follow the hierarchy $J_0 > J_1>J_2$, and we use darker shading and thicker lines to depict stronger couplings. 

We begin with the pentagonal tiling shown in Fig.~\ref{fig:Ge}(b), which realizes a so-called trellis lattice (also referred to as an elongated triangular tiling). To suppress vertical hoppings between two sites, they must differ in their $z$ coordinates, e.g., with one impurity located at $(0,0,0)$ and the other at $(1/2,n,1/2)$. Along the rim of the pentagon, this suppression can be achieved by alternating the $z$ coordinates of successive sites. However, because the rim of the pentagon contains an odd number of sites, such alternation is not globally possible. Consequently, there remain certain pairs of rim sites that share the same $z$ coordinate, and for these pairs, the hopping is \textit{not} suppressed. The exchange interaction between such pairs, which we denote $J_2$, is nevertheless much weaker than the dominant nearest-neighbor exchange $J_1$, since the former geometric separation is larger. Thus, the effective spin Hamiltonian for this system takes the form of the $J_1$--$J_2$ Heisenberg model in Eq.~\eqref{eq:J1J2}. 
We diagonalize this Hamiltonian for a 16-site cluster and determine the ground-state spin as a function of $J_2/J_1$. Over a broad range of $J_2/J_1$, the cluster stabilizes a high-spin state with $S=7/2$, which corresponds to all rim spins aligning antiferromagnetically with the four hub spins. The ratio $J_2/J_1$ given by Eq.~\eqref{eq:ratio} is $\approx 3.76 \times 10^{-2}$ for the relevant geometry when the nearest-neighbor separation is $6 \,a_B^*$. This places the system deep within the $S=7/2$ phase. In the thermodynamic limit, the pentagonal tiling leads to $s \equiv S/N \rightarrow 1/10$, unlike the square tiling which leads to an $S=0$ singlet state.

Higher values of $s$ can be obtained by tiling larger polygons. Four examples involving hexagons are presented in Fig.~\ref{fig:Ge}: (c) a honeycomb lattice, (d) a square-hexagon lattice, (e) a chamfered square tiling (also called a semitruncated square tiling), and (f) a variant of the ruby lattice\footnote{Strictly speaking, the ruby lattice is a 1-uniform, or semiregular, rhombitrihexagonal $[3.4.6.4]$ tiling of the Euclidean plane whereas the lattice in Fig.~\ref{fig:Ge}(f) has a completely different vertex configuration  of $[4.6^2; 4^2.6^2; 6^3]$. However, like the ruby lattice, the latter can be assembled by arranging squares and triangles---or equivalently, nonregular hexagons---around the perimeter of a central hexagon, so in a slight abuse of terminology, and for lack of a better nomenclature, we also refer to this distorted variant as a ``ruby''.}. In our hexagonal geometry, the hub is not equidistant from all rim sites, giving rise to two distinct hub--rim exchange couplings, denoted $J_0$ and $J_1$. In addition, the nonsuppressed couplings between rim sites at equal $z$ coordinates, denoted $J_2$, remain present. By varying the vertical bond length---recall that vertical couplings are always suppressed—we can tune the ratio $J_1/J_0$: elongating the lattice vertically reduces $J_1$. To establish the magnetic ground state, we again resort to exact diagonalization. For representative parameters with $J_0 \gtrsim J_1 > J_2$, and with the simplifying assumption $J_0 = J_1$ (a conservative choice, since it underestimates the high-spin tendency), we find robust high-spin ground states. Specifically, we obtain $S$\,$=$\,$6$ for the 20-site honeycomb, $S$\,$=$\,$7$ for the 22-site square-hexagon, $S$\,$=$\,$6$ for the 20-site chamfer, and $S$\,$=$\,$15/2$ for the 25-site ruby cluster. 

\begin{figure*}[tb]
    \centering
    \includegraphics[width=\linewidth]{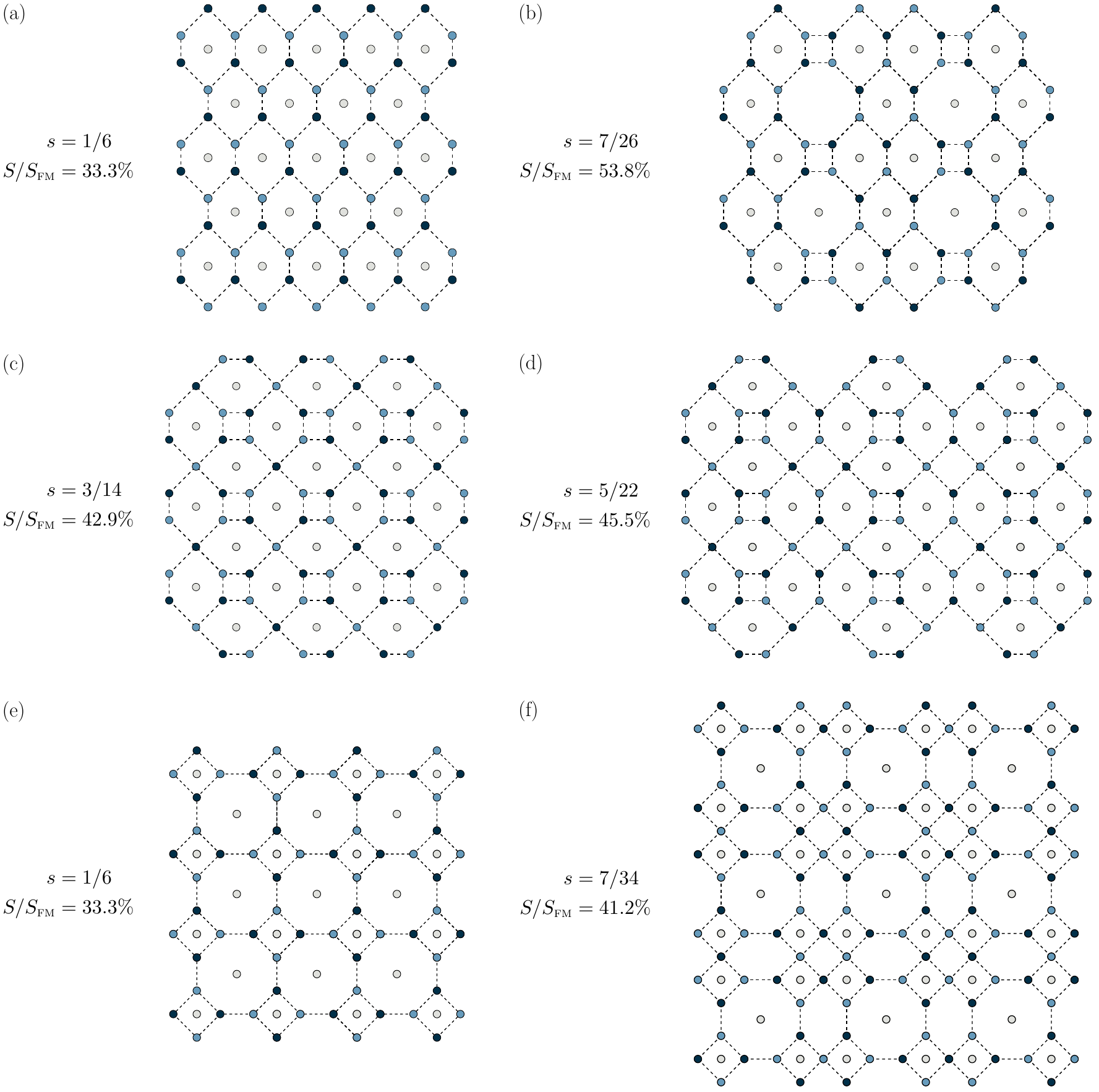}
    \caption{Different tiling possibilities for extending the individual clusters in the right column of Fig.~\ref{fig:Ge} into infinite lattices. For the four-hexagon cluster in Fig.~\ref{fig:Ge}(c), two distinct choices of translation vectors produce either (a) the honeycomb lattice or (b) the 3-uniform square-hexagon-octagon (SHO) lattice. Tessellating the chamfer in Fig.~\ref{fig:Ge}(e) forms the lattice in (c) or its expanded variant (d). Tiling the truncated-square cluster in Fig.~\ref{fig:Ge}(g) results in either (e) the square-octagon lattice or (f) the 2-uniform SHO lattice. For each construction, we indicate the spin per site $s$ in the thermodynamic limit, together with the value thereof as a fraction of the spin of the fully ferromagnetic state, $S/S_{\textsc{fm}}$.  
}
    \label{fig:octagon}
\end{figure*}

Extending these finite clusters to the thermodynamic limit requires specifying how the clusters are tiled to form an infinite lattice. Some clusters, such as in Fig.~\ref{fig:Ge}(b,d) can be replicated by simple translations and connected edge-to-edge along their outer perimeter. In contrast, the clusters tabulated in the right column of Fig.~\ref{fig:Ge} admit multiple possible tilings. In these cases, the choice of translation vector determines whether successive clusters overlap with the original unit or remain disjoint. Several representative examples of such tilings are illustrated in Fig.~\ref{fig:octagon}(a--d). Notably, the manner in which clusters are arranged has direct consequences as different tilings can yield distinct spin densities in the thermodynamic limit.  

The lattice geometries that we analyzed are summarized in Table~\ref{table:geometry}, together with their corresponding values of the net spin per site. Among these, a particularly promising example is the 3-uniform square-hexagon-octagon (SHO) lattice shown in Fig.~\ref{fig:octagon}(b), which achieves a net spin equal to $53.8\%$ of the fully polarized ferromagnetic value $S_\textsc{fm}$. In principle, there exist infinitely many $k$-uniform tilings of the plane by convex regular polygons connected edge-to-edge. Our aim here is not to exhaustively enumerate a subset of such possibilities, but rather to establish a proof of principle that high-spin states can be systematically engineered through appropriate choices of cluster tilings, and that the lattice geometry itself is a design parameter for optimizing the total spin in extended systems.

\begin{figure*}[tb]
    \centering
    \includegraphics[width=\linewidth]{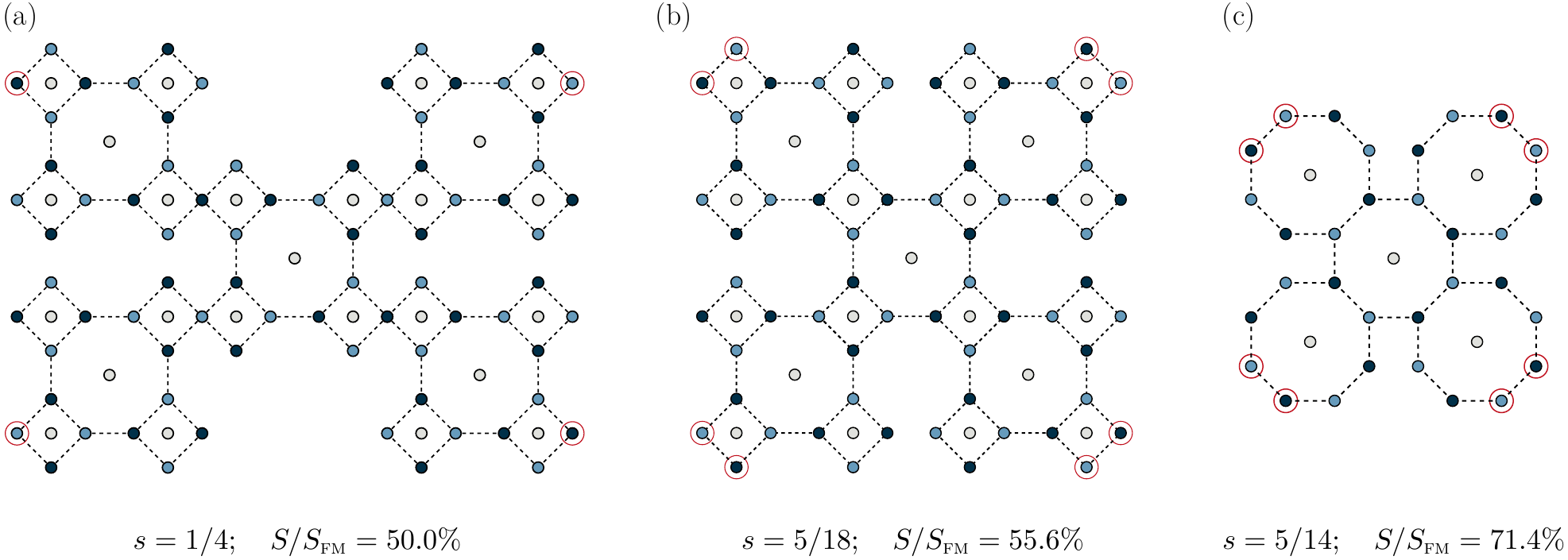}
    \caption{Three examples of generation-$g\!=\!1$ clusters illustrating different approaches to constructing fractal-like structures from (a,b) square-octagonal units and (c) octagonal units. Red circles mark the sites at which the next repeating units will be affixed to generate the subsequent iteration. Importantly, interactions between neighboring sites belonging to different $g=0$ clusters are also exponentially suppressed, allowing adjacent $g=0$ units to be positioned without adversely affecting the net spin.
}
    \label{fig:octfrac}
\end{figure*}

Finally, we consider constructing high-spin clusters using octagons as the basic building block. 
Since hoppings can only be suppressed along $45^\circ$ or $90^\circ$ angles, the octagon is the largest polygon for which a wheel-shaped cluster can be realized (i.e., we cannot engineer rotational symmetries beyond $C_4$ and $C_2$). 
As a concrete example, consider the neutral octagonal wheel shown in the central motif of Fig.~\ref{fig:Ge}(g), consisting of eight rim sites and a central hub. 
Per Eqs.~\eqref{eq:disp1} and \eqref{eq:disp2}, every other site is displaced by $(0,1/2,1/2)$ relative to the coordinates of a regular octagon. 
All nearest-neighbor hoppings along the rim are strongly suppressed, since they align with the directions listed in Eq.~\eqref{eq:cancDir}. 
The radial hub--rim hoppings are partially affected by valley degeneracy, but their amplitudes remain comparable to the hydrogenic case when the central dopant is placed at the $a_L \left(1/2, 1/2, 0\right)$ position in the unit
cell. 

Akin to some of the hexagonal tilings earlier, to tesselate the plane in a regular lattice, octagons alone are insufficient; they must be combined with squares. 
As before, enforcing alternating $z$ coordinates of successive sites ensures cancellation of vertical and horizontal hoppings, but this leaves diagonal bonds on the square plaquettes unsuppressed. 
Since these diagonal distances are comparable to the hub--rim separation in the octagon, the resulting exchanges cannot be neglected. 
To resolve this, we introduce an additional dopant at the center of each square plaquette, ensuring that each square unit is itself high-spin. 
The resulting model includes three distinct exchanges: $J_0$ on the hub--rim bonds of the square, $J_1$ on the hub--rim bonds of the octagon, and $J_2$ on the square diagonals, with $J_0\!>\!J_1\!>\!J_2$. 
For simplicity, as a  conservative estimate, we set $J_0\!=\!J_1$ and diagonalize the corresponding $J_1$--$J_2$ Heisenberg Hamiltonian on a 21-site cluster. 
For small $J_2/J_1$, the system stabilizes a high-spin ground state with $S=11/2$, in the regime dictated by our geometric construction.

Having established the stability of the octagonal unit, the natural tiling of the plane is the square--octagon lattice  (also christened the truncated square tiling, or tosquat) shown in Fig.~\ref{fig:octagon}(e). 
In the thermodynamic limit, this geometry yields $s\!\rightarrow\!1/6$. An alternative tesselation, which results in a 2-uniform SHO lattice is depicted in Fig.~\ref{fig:octagon}(f). 
Neither of these architectures  maximize the surface (or perimeter) contributions, which---as we have seen before---can significantly enhance the net spin. 
A more strategic approach therefore is to construct a fractal-like structure  by iteratively attaching octagonal wheels in the same procedure used for the square lattice. 
Starting with the unit in Fig.~\ref{fig:Ge}(g) as generation $g=0$, each subsequent generation is obtained by attaching copies of the previous generation at its four outermost vertices such that the resultant lattice is symmetric under reflections about the $x$- and $y$-axes [see Fig.~\ref{fig:octfrac}(a)]. 
For generation $g$, the total spin is 
$
S = ({8 \cdot 5^g + 3})/{2},
$
while the total number of sites is 
$
N = 16 \cdot 5^g + 5.
$
Thus, in the limit $N \rightarrow \infty$, the spin per site approaches 
$
s \rightarrow {1}/{4},
$
corresponding to $50\%$ of the magnetization of a fully polarized state. Interestingly, this coincides with the asymptotic value obtained for the square-based fractal, demonstrating that the octagonal construction provides an equally efficient route to realizing large-spin clusters. However, one can do better: even higher spin fractions can be achieved through other fractal-like geometries. Two such constructions are illustrated in Figs.~\ref{fig:octfrac}(b) and (c). The latter, obtained by generating successive iterations through gluing octagons edge-to-edge, yields a remarkably large value of 
$s = 5/14$ in the $g \rightarrow \infty$ limit, corresponding to $71.4\%$ of the spin of the  ferromagnetic state.

\setlength{\tabcolsep}{9pt}
\begin{table}[t]
\centering
\begin{tabular}{ l c l c } 
\hline
\hline
Lattice & Figure & $s$ & $S/S_{\textsc{fm}}$ \\
\hline
Square & \ref{fig:square}(a) & $0$ & --- \\
Checkerboard & \ref{fig:square}(b) & $1/6$ & 33.3\% \\
Square fractal & \ref{fig:square}(c) & $1/4$ & 50.0\% \\
Trellis   & \ref{fig:Ge}(b) & $1/10$ & 20.0\% \\
Honeycomb & \ref{fig:octagon}(a) & $1/6$ & 33.3\% \\
Square-hexagon & \ref{fig:Ge}(d) & ${1}/{4}$ & 50.0\% \\
Chamfered square  & \ref{fig:octagon}(c) & ${3}/{14}$ & 42.9\% \\
Expanded chamfer & \ref{fig:octagon}(d) & ${5}/{22}$ & 45.5\% \\
Ruby         & \ref{fig:Ge}(f) & ${5}/{22}$ & 45.5\% \\
Square-octagon  & \ref{fig:octagon}(e) & ${1}/{6}$ & 33.3\% \\
$C^{}_2$ square-octagon fractal & \ref{fig:octfrac}(a) & ${1}/{4}$ & 50.0\% \\
$C^{}_4$ square-octagon fractal & \ref{fig:octfrac}(b) & ${5}/{18}$ & 55.6\% \\
Octagon fractal & \ref{fig:octfrac}(c) & ${5}/{14}$ & 71.4\% \\
2-uniform SHO   & \ref{fig:octagon}(f) & ${7}/{34}$ & 41.2\% \\
3-uniform SHO   & \ref{fig:octagon}(b) & ${7}/{26}$ & 53.8\% \\
\hline
\hline
\end{tabular}
\caption{\label{table:geometry}The various geometric arrangements of impurity sites in Ge studied in this work. For extended lattices, we report the spin per site $s$ and the corresponding fraction of the fully polarized spin, $S/S_{\textsc{fm}}$, in the thermodynamic limit $N \rightarrow \infty$. For fractal-like arrangements, these quantities are given in the limit of generation $g \rightarrow \infty$. In both cases, finite clusters converge to their thermodynamic value of $s$ from above as $N, g \rightarrow \infty$.  
}
\end{table}

\subsection{Silicon}

Achieving strong suppression of the hopping amplitude in silicon requires a somewhat different set of design principles  than in other semiconductors. Like germanium, silicon crystallizes in a diamond cubic lattice, with atomic coordinates within the unit cell given by Eq.~\eqref{eq:diamond}. Its six equivalent conduction-band minima are located along the [100], [010], and [001] axes, at positions approximately $81\%$ of the way from the $\Gamma$ point (Brillouin zone center) to the X points (zone boundary) \cite{voisin2020valley}. The corresponding wavevectors, which are also referred to as the $\Delta$ points of the Brillouin zone, can be written as
\begin{alignat}{3}
\nonumber
\mathbf{k}^{}_1 &= \frac{2\pi}{a^{}_L} (k_0, 0, 0), \,  
&&\mathbf{k}^{}_2 = \frac{2\pi}{a^{}_L} (0, k_0, 0), \,
&&\mathbf{k}^{}_3 = \frac{2\pi}{a^{}_L} (0, 0, k_0),  \\ 
\mathbf{k}^{}_4 &= \frac{2\pi}{a^{}_L} (-k_0, 0, 0), \,  
&&\mathbf{k}^{}_5 = \frac{2\pi}{a^{}_L} (0, -k_0, 0), \,
&&\mathbf{k}^{}_6 = \frac{2\pi}{a^{}_L} (0, 0, -k_0),  
\end{alignat}
where $a_L$ is the lattice constant and we take $k_0 = 9/11 \approx  0.818$. Using a  rational approximant (i.e., assuming a commensurate conduction band minimum) makes calculations easier, and only causes deviations for \textit{very} large clusters.

The hopping integral between impurities separated by a vector  $\mathbf{R}$ is given by
\begin{align}
t(\mathbf{R}) &= \frac{1}{6} \sum_{j = 1}^6 e^{i \mathbf{k}^{}_j \cdot \mathbf{R}}\, \xi^{}_j (\mathbf{R}) 
= \frac{1}{3} \sum_{j = 1}^3 \cos \!\left(\mathbf{k}^{}_j \cdot \mathbf{R} \right)\xi^{}_j (\mathbf{R}),
\end{align}
where, in the second equality, we have used the fact that $\lvert \mathbf{k}_j \cdot \mathbf{R}\rvert = \lvert \mathbf{k}_{j+3} \cdot \mathbf{R}\rvert$ ensures $\xi_j (\mathbf{R}) = \xi_{j+3} (\mathbf{R})$, as argued above. Written out explicitly, for $\mathbf{R} \equiv (R_x,R_y,R_z)$, we have
\begin{alignat}{1} \nonumber t(\mathbf{R}) = \frac{1}{3}\bigg[&\cos\left(\frac{2\pi}{a^{}_L} k^{}_0 R^{}_x\right)\xi^{}_1 (\mathbf{R}) + \cos\left(\frac{2\pi}{a^{}_L} k^{}_0 R^{}_y\right) \xi^{}_2 (\mathbf{R}) \\ + &\cos\left(\frac{2\pi}{a^{}_L} k^{}_0 R^{}_z\right)\xi^{}_3 (\mathbf{R}) \bigg]. \label{eq:3D} \end{alignat}

From the structure of Eq.~\eqref{eq:3D}, it is clear that cancellation \textit{between} different cosine terms is not possible: the condition $\lvert k_0 R_\mu \rvert \neq \lvert k_0 R_\nu \rvert$ (as required for any two of the cosines to have equal and opposite values) implies that $\xi_\mu (\mathbf{R}) \neq \xi_\nu (\mathbf{R})$. Thus, suppression of $t(\mathbf{R})$ requires each cosine factor to vanish individually. 

Impurities can of course only occupy discrete lattice sites, so $\mathbf{R}$ is restricted to integer multiples of $a_L$ plus small intracell displacements.  Since the $\Delta$ valleys lie in the interior of the Brillouin zone, the inner products $\mathbf{k}_j \cdot \mathbf{R}$ are generally irrational multiples of $\pi$. To enforce cancellation, $\mathbf{R}$ must therefore be chosen precisely with $R_\mu = (2p+1)/(4k_0)$ for $p \in \mathbb{Z}$. Taking $k_0 = 9/11$, one convenient choice is $p=40$, giving $R_\mu = R_0 = 24.75a_L$. Using the silicon lattice constant $a_L = 5.43$A, this corresponds to an intersite distance of $134.4$A, about $7.5$ times the effective Bohr radius ($\approx 18$A).

This process can now be systematized to construct extended impurity clusters with large spin. For instance, consider a rhombus formed with dopants positioned at $(0,0,0)$, $(R_0,R_0,R_0)$, $(R_0,R_0,-R_0)$, $(2R_0, 2R_0,0)$, and at the center $(R_0,R_0,0)$. This rhombus lies in the plane defined by $x$\,$-$\,$y$\,$=$\,$0$, so practically speaking, one can implant dopants on the two-dimensional surface by simply growing the silicon crystal in the $[1\bar{1}0]$ direction. Along the rhombic edges, the hopping amplitudes cancel, leaving dominant exchange interactions that antialign the four rim spins with the central one, producing a ground state with total spin $S=3/2$. Given that the rhombus is four-coordinated like the square lattice, it admits the same tilings as shown in Fig.~\ref{fig:square}, both in uniform planar arrangements and in hierarchical, fractal-like constructions, just with the square units suitably distorted.

In bulk silicon, the six conduction-band minima are equivalent, reflecting the cubic symmetry of the lattice. The application of strain, however, breaks this symmetry and lifts the valley degeneracy; the energies of the valleys shift in a manner determined by both the strain tensor and the orientations of the valley wavevectors. To first order in the strain, the energy shift of the conduction-band minimum associated with a valley oriented along unit vector $\hat{\mathbf{k}}_j$ is described by the deformation potential
\begin{equation}
\label{eq:deform}
\Delta E_j \;=\; \Xi_d \,\mathrm{Tr}(\varepsilon) \;+\; \Xi_u \,\left(\hat{\mathbf{k}}_j\!\cdot\!\varepsilon\!\cdot\!\hat{\mathbf{k}}_j\right),
\end{equation}
where $\varepsilon$ is the symmetric strain tensor, $\Xi_d$ is the dilatation (hydrostatic) deformation potential, and $\Xi_u$ is the uniaxial deformation potential \cite{herring1956transport}. The first term produces a uniform hydrostatic shift of all valleys, while the second term is valley-dependent and is responsible for lifting the degeneracy. 

A particularly important case is biaxial strain in, say, the $(001)$ plane, which arises, for example, when a Si layer is grown epitaxially on a relaxed Si$_{1-x}$Ge$_x$ substrate \cite{yu2008first}. For an in-plane biaxial strain, $\varepsilon_{xx}\!=\!\varepsilon_{yy}\!=\!\varepsilon_\parallel$ is related to the out-of-plane component $\varepsilon_{zz}$ by Poisson’s ratio. Under these conditions, the two valleys whose axis is normal to the plane, $(0,0,\pm k_0)$ (labeled $\Delta_2$), experience a different projection $\hat{\mathbf{k}} \cdot \varepsilon \cdot \hat{\mathbf{k}}$ than the four in-plane valleys, $(\pm k_0,0,0)$ and $(0,\pm k_0,0)$ (denoted $\Delta_4$). 

For biaxial \emph{tensile} strain in the $(001)$ plane ($\varepsilon_\parallel>0$), the out-of-plane component $\varepsilon_{zz}$ is negative due to Poisson contraction. As a result, the $\Delta_2$ valleys shift downward relative to the $\Delta_4$ valleys (for the usual positive sign of $\Xi_u$ in silicon). This splits the sixfold degeneracy  into a lower twofold set ($\Delta_2$) and an upper fourfold set ($\Delta_4$) \cite{ghosh2016modeling,voisin2022valley}. Electrons preferentially occupy the $\Delta_2$ valleys, giving an impurity ground-state wavefunction of the form
\begin{equation}
\Psi(\r) = \sqrt{\frac{1}{2}} \sum_{j = 3,6} F^{}_j(\r)\, u^{}_j(\r)\, e^{i \mathbf{k}_j \cdot \r},
\end{equation}
with an associated hopping integral
\begin{alignat}{1} 
\label{eq:t1D}
t(\mathbf{R}) = \bigg[\cos\!\left(\frac{2\pi}{a^{}_L} k^{}_0 R^{}_z\right)\,\xi^{}_3 (\mathbf{R}) \bigg].
\end{alignat}

This simplification has remarkable practical utility. When dopants are implanted in the $x$--$z$ plane (assuming crystal growth along $[010]$), the hopping amplitude becomes independent of the $x$ coordinate and can therefore be suppressed solely through a judicious choice of $z$ positions alone. This allows the design of high-spin clusters with geometries that were inaccessible earlier. In particular, polygons with more than eight vertices can now be employed to tile the plane or to construct fractal-like high-spin structures, since the $x$-independence of the hopping relaxes the previous restriction to $45^\circ$/$90^\circ$ bond angles. Even for clusters with eight or fewer sites, this modification enables new lattice realizations. A good example is the Lieb lattice, a decorated square lattice with an additional site placed on each edge of the square \cite{PhysRevLett.62.1201}. Its unit cell contains three sites: two with coordination number 2 and one with coordination number 4. In the unstrained case (or for any of the other host semiconductors studied above), large-spin states could not be realized on the Lieb lattice. This is because placing a hub spin at the center of a square plaquette produced a coupling to the 2-coordinated sites that was no stronger than the coupling between adjacent 2-coordinated sites, which favors antiferromagnetism. With the present simplification, however, this problematic diagonal coupling can be eliminated just by an appropriate choice of the vertical separation of the 2-coordinated sites, thereby stabilizing high-spin ground states.

For biaxial \emph{compressive} strain in the $(001)$ plane ($\varepsilon_\parallel<0$), the sign of $\varepsilon_{zz}$ reverses, and the ordering of the valley energies flips. In this case, the four in-plane valleys ($\Delta_4$) lie lower in energy than the two out-of-plane valleys ($\Delta_2$) \cite{windbacher2010engineering,Sverdlov2011}. The impurity wavefunction is then
\begin{equation}
\Psi(\r) = \frac{1}{2}\sum_{j = 1,2,4,5} F^{}_j(\r)\, u^{}_j(\r)\, e^{i \mathbf{k}_j \cdot \r},
\end{equation}
and the hopping integral takes the form
\begin{alignat}{1} 
t(\mathbf{R}) = \bigg[\cos\!\left(\frac{2\pi}{a^{}_L} k^{}_0 R^{}_x\right)\xi^{}_1 (\mathbf{R})
+\cos\!\left(\frac{2\pi}{a^{}_L} k^{}_0 R^{}_y\right)\xi^{}_2 (\mathbf{R})\bigg].
\end{alignat}
In this regime, we can place impurities in the $(001)$ plane to form high-spin clusters. For example, a familiar wheel-shaped square cluster defined by dopants at $(0,0,0)$, $(R_0,R_0,0)$, $(R_0,-R_0,0)$, and $(2R_0,0,0)$, together with a central dopant at $(R_0,0,0)$, satisfies the cancellation conditions for hoppings along the square edges. The remaining exchange interactions on the spokes bring the cluster into a ground state with $S=3/2$. Such square units can be used to build extended structures in the two-dimensional plane in a manner, analogous to the constructions shown for AlAs in Fig.~\ref{fig:square}.

For realistic engineering strains (from fractions of a percent up to a few percent), the valley splittings typically fall in the meV range \cite{boykin2004valley,PhysRevB.108.125405}.
We note in passing that other forms of strain can also lead to similar valley splitting. For instance, uniaxial strain along $[001]$ separates the six valleys into two groups: the pair aligned with the strain axis and the four orthogonal valleys \cite{PhysRev.132.1080,ungersboeck2008effect}. Their relative ordering depends on the sign of the applied strain and on the sign of $\Xi_u$. In silicon, for tensile uniaxial strain and positive $\Xi_u$, the valleys along the strain axis ($\Delta_2$) are lowered relative to the others. More generally, strain along non-high-symmetry directions (e.g., $[110]$ or shear) can mix valley character and produce more complicated couplings and splittings into two or three distinct groups depending on the microscopic details.

\section{Discussion and outlook}
\label{sec:end}

\begin{figure*}[tb]
\includegraphics[width=\linewidth]{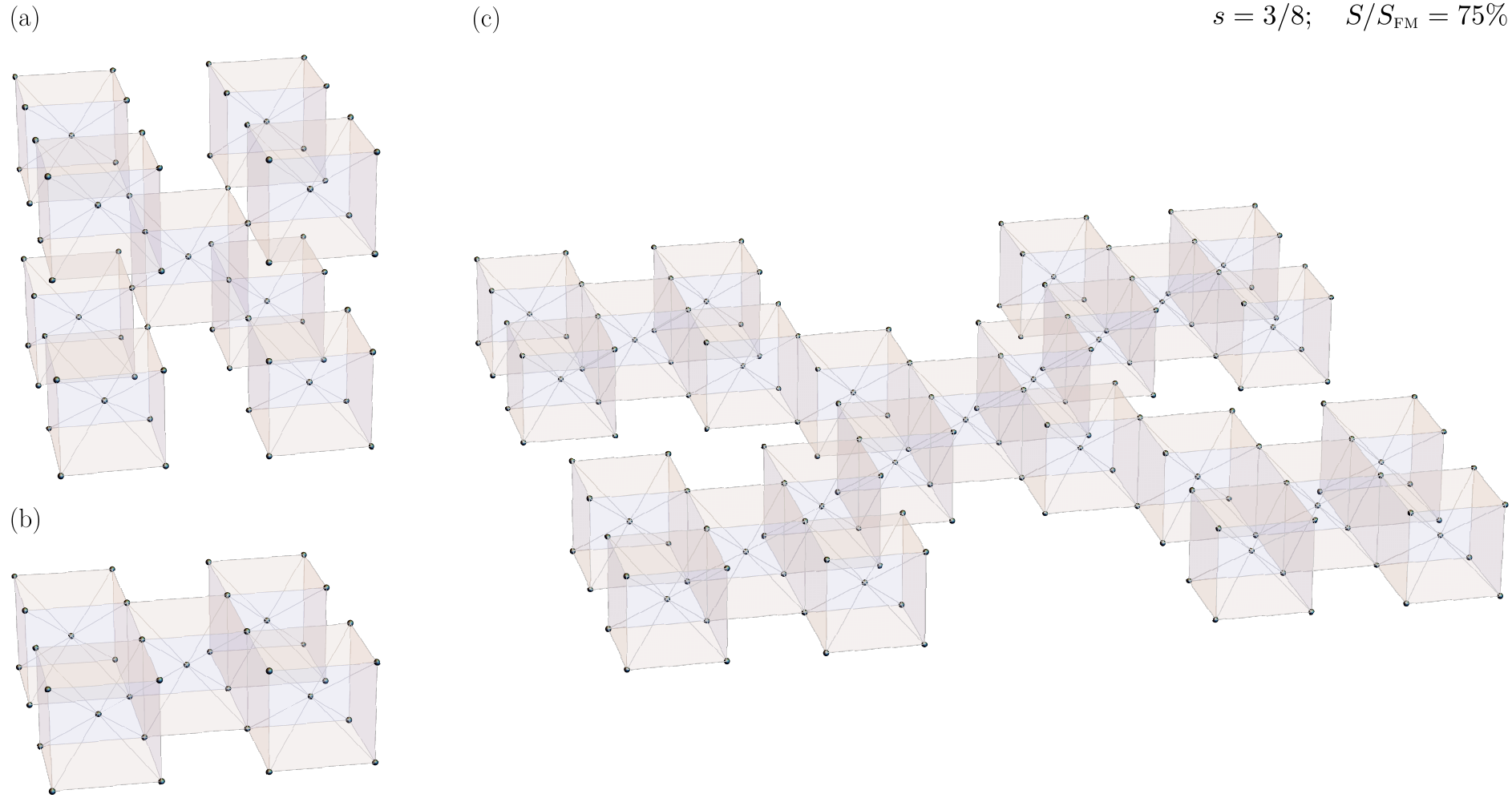}
\caption{Fractal-like high-spin structures generated by iterating a cubic building block. Each cube consists of nine dopants arranged in a body-centered cubic geometry, with eight impurities at the vertices and one central ``hub'' site. For GaAs hosts, such an elementary cube exhibits a high-spin ground state due to the same geometric considerations as in two dimensions [see Fig.~\ref{fig:wheel}(b)]. In AlAs or Ge, the couplings along the edges of the cube can further be exponentially suppressed. (a) A fully three-dimensional and (b) a quasi-two-dimensional generation-$g=1$ cluster constructed from these cubes. Higher generations are obtained by attaching additional cubes to the outermost vertices of the existing structure: eight in 3D and four in the quasi-2D case. (c) Quasi-2D generation-$g=2$ structure realized through this iterative process. The net spin scales as $(6\cdot 5^g+1)/2$, while the number of sites grows as $(8\cdot 5^g+1)$. Thus, in the limit $g \rightarrow \infty$, the spin per site approaches $s \rightarrow 3/8$, corresponding to $75\%$ of the fully polarized value.
}
\label{fig:3D}
\end{figure*}

In this work, we have developed a systematic method for generating high-spin clusters using substitutional impurities in semiconductors, achieving net spin values that scale extensively with the system size. Our approach focused on uncompensated semiconductors with degenerate and anisotropic conduction band minima, where impurity wavefunctions and energies deviate from the simple hydrogenic form, rendering the magnetic behavior far richer than in single-valley systems.  Within the tight-binding approximation, we derived a general expression for the hopping parameter in multivalley semiconductors. This analysis revealed the possibility of suppressing hopping amplitudes between selected impurity sites, a feature that can be exploited to stabilize high-spin ground states. Using this principle, we then constructed a variety of cluster geometries, ranging from isolated structures to extended lattices and fractal-like arrangements. Representative results for these configurations are summarized in Table~\ref{table:geometry}.

Among the lattice-based designs, we highlight one particularly favorable tiling: the 3-uniform SHO lattice [Fig.~\ref{fig:octagon}(b)], which is realizable through dopant implantation in Ge. In the thermodynamic limit, this structure supports a net spin equal to $53.8\%$ of the fully polarized value, which is the largest fraction we have identified for uniform lattices in two dimensions. For fractal-like structures, the largest spin among the geometries studied here is obtained for the octagonal cluster [Fig.~\ref{fig:octfrac}(c)], which yields $71.4\%$ of the ferromagnetic value.
Although this is the best result within the families we considered, the space of possible lattices is infinite, and more favorable designs may exist.

As a general principle, we find that fractal arrangements maximize perimeter contributions to the net spin and therefore yield higher spin fractions. In two dimensions, tilings based on squares and octagons approach a limiting value of $S/S_{\textsc{fm}}=50\%$ and $S/S_{\textsc{fm}}=75\%$, respectively, as the generation $g\rightarrow\infty$. Strained Si, however, allows for the construction of fractal patterns based on polygons with larger numbers of vertices. With this generalization, one can construct fractal-like patterns using any $m$-sided polygon with $C_4$ rotational symmetry (i.e., $m$ divisible by four). In this case, assuming one overlapping impurity vertex between each pair of adjacent units, the asymptotic spin per site follows
$
s \rightarrow 1/2 - 1/m
$,
yielding $S/S_{\textsc{fm}}=75.0\%$, $83.3\%$, and $87.5\%$ for $m=8$, $12$, and $16$, respectively.  

The theoretical framework presented here naturally generalizes to three spatial dimensions. An example is shown in Fig.~\ref{fig:3D}(a), for which the host lattice may equally well be GaAs, AlAs, or Ge. The elemental unit for this structure is a three-dimensional analogue of the wheel-shaped cluster: a cube with a central dopant atom. If the hoppings along the cube's edges are canceled out or geometrically suppressed, the central site enforces alignment of all the vertex spins, resulting in a high-spin ground state. Repetition of this structure in a fractal pattern leads to a net spin per site of $s=3/8$, or $S/S_{\textsc{fm}}=75\%$, which exceeds that of all planar clusters examined in this work.  
Although precise three-dimensional dopant positioning remains experimentally challenging, quasi-two-dimensional realizations may be more accessible and are just as useful. For instance, constructing a single-layer array of such cubic units [Figs.~\ref{fig:3D}(b,c)] produces a high-spin ground state with $S/S_{\textsc{fm}}\!=\!71.4\%$, which is reasonably close to the full 3D value.

In order to realize large-spin magnetic states, most of the clusters considered here have to be constructed with sufficiently large interdopant separations ($\sim5$--$6\,a^{*}_B$) so as to ensure that the electrons do not become itinerant. Unfortunately, this implies that the associated energy splittings are typically very small, making such clusters highly susceptible to thermal effects. This follows from the fact that the effective Rydberg of shallow impurity centers is much smaller than the hydrogenic Rydberg, thereby setting a correspondingly low energy scale for the bound states of shallow impurity clusters. One natural extension of this work therefore is to explore the properties of clusters formed from \textit{deep} impurity sites, where the relevant energy scales are significantly larger. In this regime, however, donor wavefunctions become strongly localized near the impurity, and the short-range atomic potential plays a dominant role. As a result, the wavefunctions and bound-state energies deviate substantially from hydrogenic forms, requiring a more microscopic theoretical description. Furthermore, because of this strong localization, clusters are confined to volumes comparable to the host lattice constant, and the discreteness of the underlying lattice must be taken into account in their design.  

With regard to the use of semiconductors with degenerate conduction-band minima as hosts for dopant clusters, our work also pinpoints several important questions for future study. A more quantitative analysis of such non-hydrogenic impurities will have to be carried out within an atomistic framework, such as quantum chemistry or first-principles calculations. While we illustrated neutral clusters that exploit cancellations of specific hopping amplitudes, the analogous possibilities for charged clusters remain largely unexplored. Previous work has shown that negatively charged clusters can be stabilized in multivalley semiconductors at small separations \cite{Bhatt}. It would be worthwhile to investigate whether this stabilization persists at separations on the order of multiple effective Bohr radii, especially since prior studies indicate that negatively charged clusters are good candidates for realizing high-spin ground states \cite{Nielsen}.  

An additional overarching challenge arises from the sensitivity of hopping amplitudes to the precise placement of impurity sites. For hydrogenic clusters, small errors in dopant positions produce only modest deviations, as the effective parameters vary smoothly with separation. By contrast, in semiconductors with degenerate band minima, hopping integrals can change sharply with small displacements of impurity sites within the unit cell. This prompts a systematic search for cluster geometries that are inherently robust against such placement errors.

It is also worth noting that the cancellations in hopping amplitudes that we exploited here were derived within the tight-binding approximation. However, we emphasize that the suppression due to interference between different valleys is much more robust than the effective hopping within a tight-binding model. In fact, the same effect is obtained with a Heitler-London approximation for donor pairs using the full Coulomb interaction \cite{Andres, koiller2001exchange,koiller2002strain}, as well as in more microscopic approaches \cite{wellard2003,voisin2020valley}. Although exact cancellations may not persist in more refined treatments going beyond the effective mass approximation, our results nevertheless highlight promising routes for stabilizing high-spin clusters in multivalley semiconductors.

Taken together, our findings motivate experimental efforts to better characterize impurity clusters in materials such as AlAs and Ge, where the underlying valley structure is particularly favorable.  For example, the binding energy of a single shallow impurity state can deviate measurably from one Ry$^*$ due to short-range corrections, and empirical investigations of such effects would be useful for developing a quantitatively accurate picture. 
Finally, the magnetic properties of impurity clusters calculated here, even if approximate, provide concrete predictions for future experiments. Our proposal is readily implementable with today's device technologies \cite{Simmons} and  can thus be directly tested and refined through large-scale quantum simulation experiments using, say, gate-defined semiconductor  quantum dot arrays \cite{Vandersypen, wang2022experimental, kiczynski2022engineering}.

\section*{Acknowledgments}
We thank Andrew Osborne for helpful discussions. 
This work was performed in part at the Aspen Center for Physics, which is supported by a grant from the Simons Foundation (1161654, Troyer). R.S. was supported by the Princeton Quantum Initiative Fellowship. The numerical calculations presented in this article were performed on  computational resources managed and supported by Princeton Research Computing, a consortium of groups including the Princeton Institute for Computational Science and Engineering (PICSciE) and the Office of Information Technology's High Performance Computing Center and Visualization Laboratory at Princeton University.

\bibliography{refs}

\end{document}